\documentclass[twocolumn, tighten]{aastex61}
\usepackage{graphicx}
\begin{document}
\title{The Ages of Passive Galaxies in a $z$=1.62 Protocluster}
\shorttitle{The Ages of Passive Galaxies in a $z$=1.62 Protocluster}
\shortauthors{Lee-Brown, et al.}
\author[0000-0003-0164-0600]{Donald B. Lee-Brown}
\affiliation{Department of Physics and Astronomy, University of Kansas, Lawrence, KS 66047, USA}
\correspondingauthor{Donald Lee-Brown}
\email{donald@ku.edu}
\author[0000-0001-5851-1856]{Gregory H. Rudnick}
\affiliation{Department of Physics and Astronomy, University of Kansas, Lawrence, KS 66047, USA}
\author{Ivelina G. Momcheva}
\affiliation{Space Telescope Science Institute, 3700 San Martin Drive, Baltimore, MD 21218, USA}
\author[0000-0001-7503-8482]{Casey Papovich}
\affiliation{Department of Physics and Astronomy, Texas A\&M University, College
Station, TX, 77843-4242 USA}
\affiliation{George P.\ and Cynthia Woods Mitchell Institute for
  Fundamental Physics and Astronomy, Texas A\&M University, College
  Station, TX, 77843-4242}
\author{Jennifer M. Lotz}
\affiliation{Space Telescope Science Institute, 3700 San Martin Drive, Baltimore, MD 21218, USA}
\author[0000-0001-9208-2143]{Kim-Vy H. Tran}
\affiliation{Department of Physics and Astronomy, Texas A\&M University, College
Station, TX, 77843-4242 USA}
\affiliation{George P.\ and Cynthia Woods Mitchell Institute for
  Fundamental Physics and Astronomy, Texas A\&M University, College
  Station, TX, 77843-4242}
\author{Brittany Henke}
\affiliation{Department of Physics and Astronomy, University of Kansas, Lawrence, KS 66047, USA}
\author{Christopher N. A. Willmer}
\affiliation{Steward Observatory, University of Arizona, 933 N. Cherry Avenue, Tucson, AZ 85721, USA}
\author[0000-0003-2680-005X]{Gabriel B. Brammer}
\affiliation{Space Telescope Science Institute, 3700 San Martin Drive, Baltimore, MD 21218, USA}
\author{Mark Brodwin}
\affiliation{Department of Physics and Astronomy, University of Missouri, 5110 Rockhill Road, Kansas City, MO 64110, USA}
\author{James Dunlop}
\affiliation{Institute for Astronomy, University of Edinburgh, Royal Observatory, Edinburgh, EH9 3HJ, UK}
\author{Duncan Farrah}
\affiliation{Department of Physics, Virginia Tech, Blacksburg, VA 24061, USA}

\begin{abstract}
We present a study of the relation between galaxy stellar age and mass for 14 members of the $z=1.62$ protocluster IRC 0218, using multiband imaging and HST G102 and G141 grism spectroscopy. Using $UVJ$ colors to separate galaxies into star forming and quiescent populations, we find that at stellar masses $M_* \geq 10^{10.85} M_{\odot}$, the quiescent fraction in the protocluster is $f_Q=1.0^{+0.00}_{-0.37}$, consistent with a $\sim 2\times $ enhancement relative to the field value, $f_Q=0.45^{+0.03}_{-0.03}$. At masses $10^{10.2} M_{\odot} \leq M_* \leq 10^{10.85} M_{\odot}$, $f_Q$ in the cluster is $f_Q=0.40^{+0.20}_{-0.18}$, consistent with the field value of $f_Q=0.28^{+0.02}_{-0.02}$. Using galaxy $D_{n}(4000)$ values derived from the G102 spectroscopy, we find no relation between galaxy stellar age and mass. These results may reflect the impact of merger-driven mass redistribution, which is plausible as this cluster is known to host many dry mergers. Alternately, they may imply that the trend in $f_Q$ in IRC 0218 was imprinted over a short timescale in the protocluster's assembly history. Comparing our results with those of other high-redshift studies and studies of clusters at $z\sim 1$, we determine that our observed relation between $f_Q$ and stellar mass only mildly evolves between $z\sim 1.6$ and $z \sim 1$, and only at stellar masses $M_* \leq 10^{10.85} M_{\odot}$. Both the $z\sim 1$ and $z\sim 1.6$ results are in agreement that the red sequence in dense environments was already populated at high redshift, $z\ga 3$, placing constraints on the mechanism(s) responsible for quenching in dense environments at $z\ge 1.5$.
\end{abstract}


\section{Introduction}
In the local Universe, galaxies populate two distinct regions in color-magnitude space. Red, massive galaxies with predominantly early-type morphologies and little or no ongoing star formation form a tight relation known as the \emph{red sequence}, while the \emph{blue cloud} consists of lower-mass, star forming galaxies with late-type morphologies \citep{Kauffmann03}. This bimodality is unambiguously observed to redshifts $z>2$ (e.g.~\citealt[][]{Kriek08, Williams09, Whitaker13}), and massive quiescent galaxies have been detected as early as $z\sim 4$ \citep{Straatman14}. 

The properties of galaxies also depend on their environment. Numerous studies have found an inverse relation between average specific star formation rate (sSFR) and galaxy overdensity, such that the average sSFR in cluster cores is much lower than that in the field, at least at low redshifts (e.g. \citealt{Hashimoto98, Lewis02, Gomez03, Hogg04, Balogh04, Patel09, Chung11}). Additionally, by $z\sim 0$, the fraction of quiescent, lower-mass cluster galaxies relative to the field is significantly enhanced and approaches the value for the more massive central galaxies \citep{Tal14}. These results indicate that in addition to the intrinsic mechanisms responsible for the quenching of massive galaxies \citep[e.g.,][]{Croton06, Peng10}, environmentally-driven processes exist and may even play a dominant role in the quenching of lower-mass galaxies (e.g. \citealt{Bassett13, Delaye14}). A comprehensive understanding of the physical processes responsible for quiescence and their relative importance as a function of galaxy mass and environment is one of the primary objectives in extragalactic astrophysics.

Attempts to observe young clusters that are in the process of virializing have naturally pushed to higher redshifts, where massive ($M_* \ge 10^{10} M_{\odot}$) quiescent galaxies are building up the bulk of their stellar populations and actively shutting off their star formation. At $z\ga 1$, the mix of galaxy populations is significantly different than that observed at lower redshifts, with some clusters showing evidence for high fractions of star forming galaxies (e.g. \citealt{Butcher78, Tadaki12, Brodwin13}) compared with lower-redshift clusters, recent or ongoing merging \citep{Rudnick12,Lotz13}, and star forming early-type galaxies \citep{Mei15}. Additionally, studies have argued that environmental quenching mechanisms play a secondary role in suppressing star formation rate at these high redshifts, with intrinsic processes responsible for the bulk of the observed red sequence galaxies  \citep[e.g.,][]{Quadri12, Brodwin13, Fassbender14, Nantais16}. 

Still, even at $z \ga 1.5$, the environment likely still plays some role in cluster red sequence assembly. In recent years a steadily increasing number of high-redshift ($z>1.5$) clusters or protoclusters have been discovered and confirmed (e.g. \citealt{Andreon09, Papovich10, Gobat11, Santos11, Stanford12, Zeimann12, Muzzin13b, Mei15, Webb15, Cooke16, Wang16}). What studies have found is that while significant cluster-to-cluster scatter exists, $z\ga 1.5$ clusters often still have well-developed red sequences and elevated quiescent fractions relative to the field (\citealt{Quadri12, Newman14, Cooke16, Nantais16}, though see \citealt{Lee15}). However, it is not yet clear what mechanisms drive the observed enhanced quiescent fractions. 

To address this deficiency, we focus in this work on one high-redshift dense environment, the $z=1.62$ protocluster XMM-LSS J02182-05102 (also known as IRC 0218 and hereafter referred to as such). The protocluster was identified as an overdensity in red \textit{Spitzer}/IRAC colors coincident with weak \textit{XMM} X-ray emission, and subsequently spectroscopically confirmed with IMACS on the Magellan Telescope \citep{Papovich10}. It was simultaneously and independently discovered and confirmed through similar methods by \citet{Tanaka10}. Various followup studies have built up a wealth of multiband photometric and spectroscopic observations (see Section~\ref{Sec:data}) making the protocluster an important object for the study of high-redshift dense environments. 

The mass of IRC 0218 is estimated to be around $M \sim 4-7 \times 10^{13} M_{\odot}$ \citep{Papovich10, Finoguenov10, Pierre12, Tran15}, which is lower than the most massive clusters observed at $z>1.5$ \citep[see, e.g.,][]{Stanford12, Brodwin16}. Its modest mass, coupled with significant expected future accretion, make it a likely progenitor of a Virgo-mass cluster in the local Universe \citep{Hatch16}. The weakness of its intracluster X-ray emission as observed by \textit{Chandra} indicates that it is not yet virialized \citep{Pierre12}, with recent work supporting this and finding that the protocluster actually consists of several distinct subgroups in the process of assembly \citep{Hatch16}. 

The protocluster's unevolved state and low galaxy-galaxy velocity dispersion suggest a much higher merger rate than in more mature cluster environments \citep{Papovich10, Pierre12, Rudnick12, Hatch16}. Indeed, morphological studies of the red sequence population have found a lack of compact quiescent galaxies, attributed to a previous and/or currently increased merger rate in the protocluster relative to the field \citep{Papovich12}. This explanation is supported by the direct-imaging study of \citet{Lotz13}, which found that the merger rate for the massive galaxies ($M_* \ge 10^{10} M_{\odot}$) is a factor of 3-10 higher than that in the field, with the bulk of the extra merging due to dry minor merging. More recently, the brightest cluster galaxy was shown to be a record-breaking gravitational lens rather than a merger \citep{Wong14}, but even after adjusting for this IRC 0218 still shows an elevated merger rate relative to the field. 

Since IRC 0218 has a relatively well-developed red sequence, is still in the process of assembly, and shows evidence for elevated merger activity, it is an excellent target for investigating the buildup of the high-redshift red sequence in dense environments. In this work, we examine the relation between galaxy mass and $D_{n}(4000)$ (which we use as an indicator of stellar age) in order to constrain the possible explanations for the buildup of the red sequence in dense environments. For example, if mass is the primary driver of quenching at high redshifts, then there should be a clear relation between age and stellar mass. On the other hand, if, e.g., mergers have played a significant role in the buildup of the cluster's red sequence, we expect little to no relation between mass and age on the red sequence. Other environmental processes that work over a range in masses such as strangulation \citep{Larson80} may also serve to flatten any $D_{n}(4000) - M_*$ relation, though the low mass of the protocluster means that other possible mechanisms such as vigorous ram pressure stripping effects \citep{Gunn72} are unlikely to produce such an effect. 

The outline of this work is as follows. Our data is described in Section~\ref{Sec:data}. In Section~\ref{Sec:sample} we explain how we selected our cluster and field samples, measured stellar masses and the $D_{n}(4000)$ spectroscopic index \citep{Bruzual83, Balogh99}, and estimated sample completeness. In Section~\ref{Sec:results} we discuss the quiescent fraction of galaxies in the cluster and field samples, and use the measurements of $D_{n}(4000)$ to constrain age differences between quiescent and star forming cluster galaxies. Finally, we discuss our results in Section~\ref{Sec:discussion}, compare them with the results of other high-redshift galaxy cluster studies, and conclude in Section~\ref{Sec:conclusion}. Throughout, we use a $\Lambda$CDM cosmology with $\Omega_{m}=0.3$,$\Omega_{\Lambda}=0.7$, and $H_0=70$ km s$^{-1}$ Mpc$^{-1}$, and a Chabrier IMF \citep{Chabrier03}. Magnitudes given are in the AB system with zeropoint $M_{AB}=25$. 

\section{Data}
\label{Sec:data}
IRC 0218 is in the UKIDSS Ultra-Deep Survey field (UDS; \citealt{Lawrence07}) and was partially imaged with the WFC3 F125W ($J_{125}$) and F160W ($H_{160}$) filters as part of the CANDELS survey (\citealt{Grogin11, Koekemoer11}). Full CANDELS-depth photometric coverage of the cluster in F125W and F160W, as well as 10 orbits of G102 grism spectroscopy of the densest protocluster subgroup, were obtained under a followup program (PI Papovich: GO 12590). Additionally, we use WFC3 F140W filter and G141 grism coverage of the cluster collected under the 3D-HST survey \citep{Brammer12, Momcheva15}. The photometry used for this study is contained in the v4.1.5 3D-HST release, and was self-consistently reduced as described in \citet{Skelton14}, and both G102 and G141 data were reduced according to the procedure described in \citet{Momcheva15}. 

\begin{figure*}[htb!]
\epsscale{1.18}
\plotone{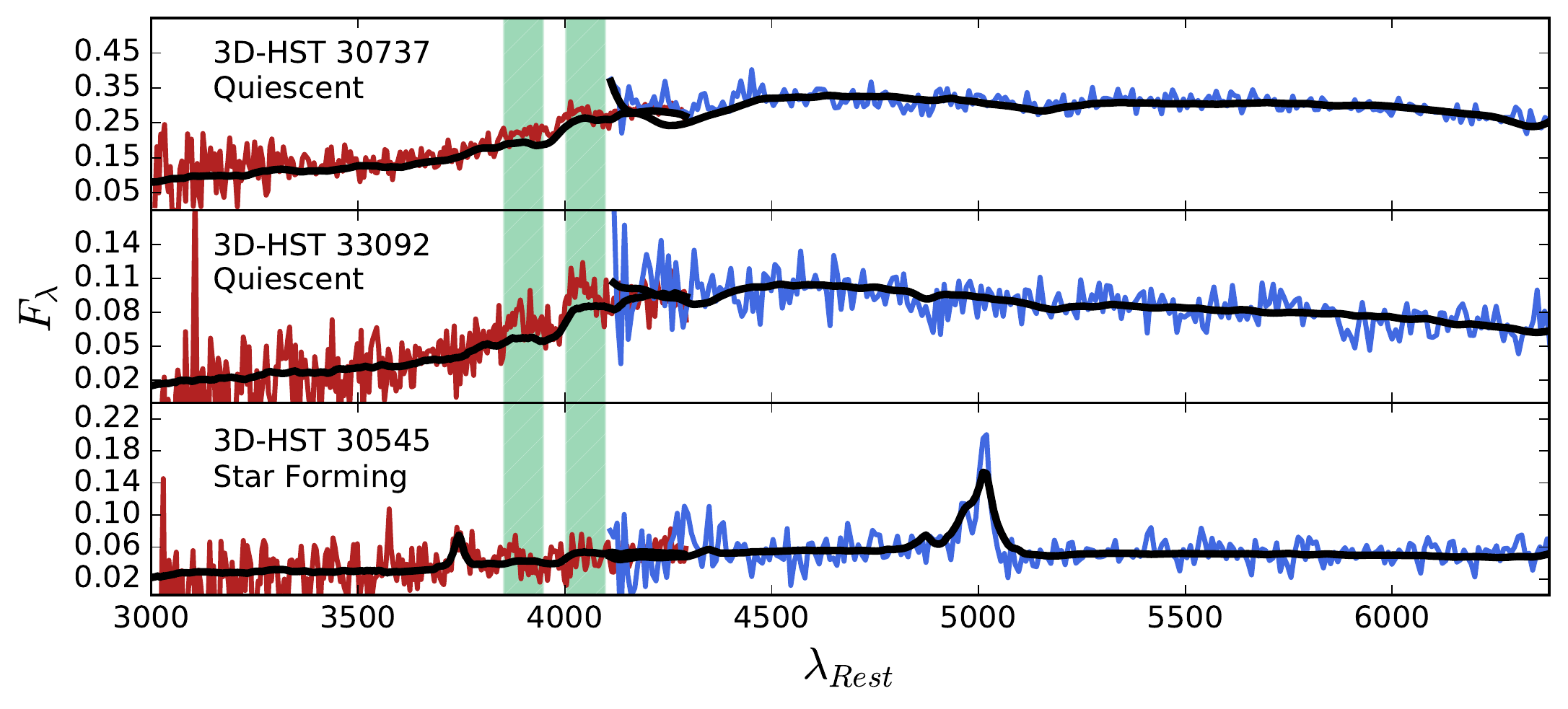}
\figcaption{Rest-frame G102 (red) and G141 (blue) grism spectra of three galaxies in IRC 0218, plotted from $\sim$ 3000-6850 $\mathrm{\AA}$. The colored lines are measured fluxes, and the black lines are best-fit SED models. The two grism spectra for each galaxy were fit independently, resulting in some model discontinuity where the grism wavelength ranges overlap, $\sim 4200$ $\mathrm{\AA}$. The vertical striped bars denote the upper and lower $D_{n}(4000) $ indices (see Section~\ref{Sec:d4kmeasure}). Galaxy classifications (quiescent or star forming) are described in Section~\ref{Sec:uvj}.}
\label{Fig:example}
\end{figure*}

\begin{figure}[htb!]
\epsscale{1.1}
\plotone{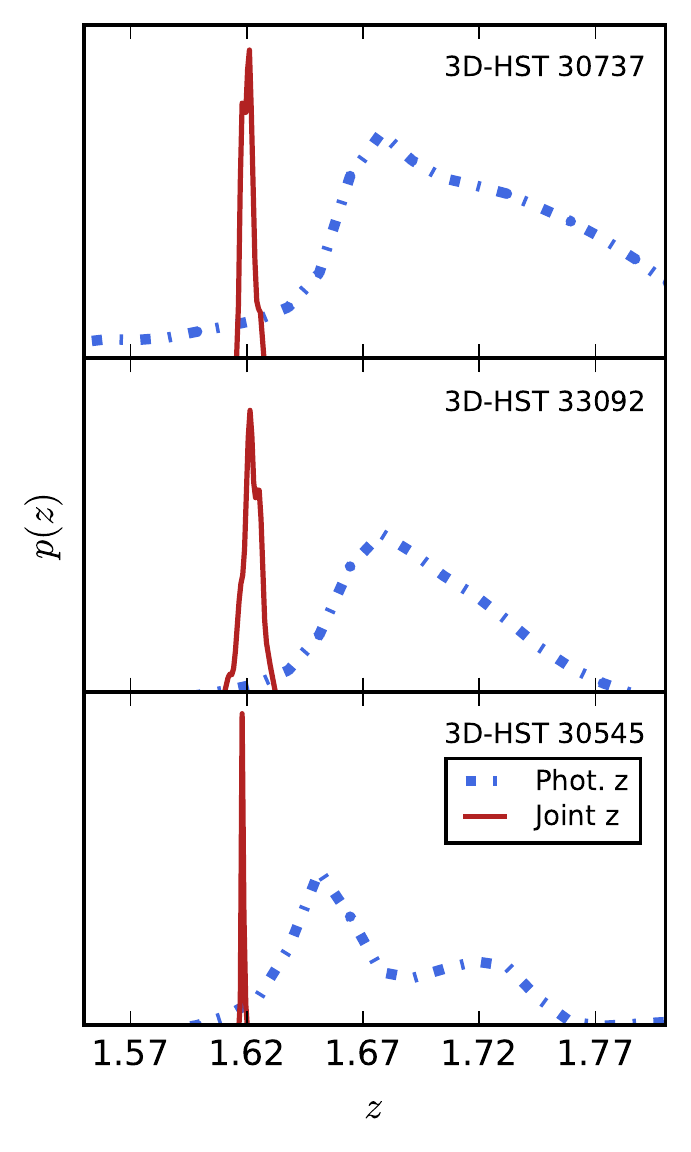}
\figcaption{Redshift probability distributions for the three galaxies shown in Figure~\ref{Fig:example}. The solid red line indicates the $p(z)$ derived from jointly-fitting the G102, G141, and photometric data. The dashed blue line shows the $p(z)$ derived from fitting the photometric data alone. For visualization purposes, the probability distributions for each galaxy are scaled by a multiplicative factor relative to one another.}
\label{Fig:pz}
\end{figure}

Several example grism spectra and their accompanying SED fits are given in Figure~\ref{Fig:example}. The G102 grism provides low-resolution ($R\sim 210$) spectroscopy of wavelengths $800\leq \lambda \leq 1150 $ nm, while the G141 grism covers $1075\leq \lambda \leq 1700 $ nm with $R\sim 130$. Thus, the grism observations cover the rest-frame $4000$ $\mathrm{\AA}$ (Balmer) break, the $3727$ $\mathrm{\AA}$ [OII] region, and the $4959$ $\mathrm{\AA}$ and $5007$ $\mathrm{\AA}$ [OIII] features at the redshift of the protocluster. The range of spectroscopic coverage provided by the two grisms, as well as the extensive photometric data, together enable precise redshift determinations through SED fitting. In Figure~\ref{Fig:pz}, we show the redshift probability distributions for the three galaxies given in Figure~\ref{Fig:example}. The probability distributions derived from the jointly-fit grism and photometric data are significantly better constrained than the distributions from photometry alone. This is largely due to resolution of fine spectral features in the grism data, and illustrates the utility of $HST$ grism data in determining galaxy redshifts at early epochs.

As a final note, some of the galaxies in IRC 0218 have redshifts derived from higher resolution ground-based spectroscopy (Subaru/MOIRCS: \citealt{Tanaka10}, Magellan/IMACS: \citealt{Papovich10}, KECK/MOSFIRE+LRIS: \citealt{Tran15}). The cases where these redshifts were adopted are indicated in Table~\ref{Table:cluster}. In general, preference was given to these redshifts except in three cases where the grism-based redshifts were indisputably superior. Nearly all of these ground-based redshifts except a few from \citet{Tanaka10} were obtained from emission lines. Significantly deeper continuum spectroscopy is needed to obtain quiescent galaxy redshifts. With the grism we are able to measure precision $4000$ $\mathrm{\AA}$ based redshifts of quiescent galaxies. The procedure by which the grism redshifts were derived is detailed in Section~\ref{Sec:d4kmeasure}. 

\section{Sample Selection and Completeness}
\label{Sec:sample}

\subsection{The Field and Protocluster Grism Samples}

We select our spectroscopic field and protocluster samples from the galaxies with G102 coverage of the rest-frame $4000$ $\mathrm{\AA}$ break, which limits our samples to galaxies in the redshift range $1.27\la z \la 1.77$. Additionally, the spatial extent of the spectroscopic samples is limited to the single 3D-HST field UDS-18, the only field for which we have the requisite G102 coverage. Comparison of the SED fit-derived grism+photometric redshifts with those obtained by higher resolution ground-based spectroscopy indicates an average deviation of $\overline{\Delta z}=0.01$ at the protocluster redshift, so we construct the cluster sample using all $25$ galaxies with redshifts $z=1.62 \pm 0.02$ in the field. Our field grism sample contains $38$ galaxies with redshifts $1.27\leq z \leq 1.58$, $1.66 \leq z \leq 1.77$, though we note that we also make use of a much larger field sample composed of photometric sources, which we detail in Section~\ref{Sec:uvj}.

\subsection{Measurement of Stellar Masses and $D_{n}(4000)$}
\label{Sec:d4kmeasure}
Stellar masses for the protocluster and field galaxies were estimated through template fits to the rest-frame photometry, using the fitting code FAST \citep{Kriek09}. For the template fits, we adopted the maximally likely redshifts from the redshift probability distributions derived from available G141, G102, and photometric data. As a galaxy's redshift probability distribution is, in general, non-Gaussian, upper and lower 68$\%$ redshift confidence intervals were estimated through resampling. The protocluster sample contains 10 galaxies with precision redshifts measured from high-resolution spectroscopy, and for these galaxies we fix their redshifts to the spectroscopic values. \citet{Bruzual03} stellar population models were used to generate templates with ages in the range $10^{7.6} \leq t/yr \leq 10^{10.1}$, and exponentially declining star formation histories with $\tau$ in the range $10^{7} \leq \tau /yr \leq 10^{10}$ yr. Metallicity was fixed at $Z=0.02$ to be consistent with the 3D-HST catalogs and to avoid the large uncertainties associated with modelling galaxy chemical evolution. The dust law of \citet{Calzetti00} was adopted for the template fitting, though we note that galaxy mass is relatively insensitive to the specific choice of dust law. 

The rest-frame $4000$ $\mathrm{\AA}$ break is due to line blanketing -- primarily Ca and Fe lines -- of the stellar continuum emission from lower-mass (F, G, and K spectral type) dwarfs and giants. Young stellar populations will have weak $4000$ $\mathrm{\AA}$ breaks due to the luminosity contribution from high-mass stars, and old populations with prominent absorption lines will have the strongest $4000$ $\mathrm{\AA}$ breaks. Thus, the $D(4000)$ spectroscopic index \citep{Bruzual83}, defined to be the ratio of flux above the $4000$ $\mathrm{\AA}$ break to the flux below it, provides a straightforward estimate of a galaxy's luminosity-weighted age (see, e.g, \citealt{Rudnick00, Kauffmann03}).  In this work, we adopt the wavelength definitions of \citet{Balogh99}, i.e., the index is defined to be the ratio of mean rest-frame flux in the region $4000 \leq \lambda \leq 4100$ $\mathrm{\AA}$ to the flux in the region $3850 \leq \lambda \leq 3950$ $\mathrm{\AA}$, and we denote this as $D_{n}(4000)$. The principal advantage of $D_{n}(4000)$ over the original index definition of \citet{Bruzual83} is its insensitivity to reddening effects, which we confirmed with extensive simulations. We note that measurements of $D_{n}(4000)$ from simulated spectra degraded to the G102 grism resolution show no systematic offsets relative to measurements at higher resolutions \citep{Henke15}.

The method used to estimate the uncertainties in $D_{n}(4000)$ is as follows. First, one-dimensional optimally weighted spectra were extracted from the 2-D G102 grism spectra following the method of \citet{Horne86}. Next, a redshift was randomly drawn from the per-galaxy redshift probability grid derived from jointly fitting the G102, G141, and photometric data. For the joint-fit redshifts, typical redshift probability grid step sizes are $\Delta z \la 10^{-3}$; we nonetheless linearly interpolated onto a $10$x denser grid before the random draw. If a galaxy had a high quality, ground-based spectroscopic redshift, this redshift was selected instead and fixed. Finally, under the assumption of normally distributed flux uncertainties, the flux in each wavelength bin was randomly drawn, and appropriately redshifted $D_{n}(4000)$ calculated. Linear interpolation was used to estimate the flux in partial wavelength bins at the index boundaries. After 100,000 iterations of this process for each galaxy, upper and lower $68\%$ confidence intervals were calculated for each measured $D_{n}(4000)$. These uncertainties are based on redshift and flux uncertainties alone; we assess the metallicity dependence of $D_{n}(4000)$ in Section~\ref{Sec:discussion}. \\ \\ \\

\subsection{Completeness}

\begin{figure}[htb!]
\epsscale{1.1}
\plotone{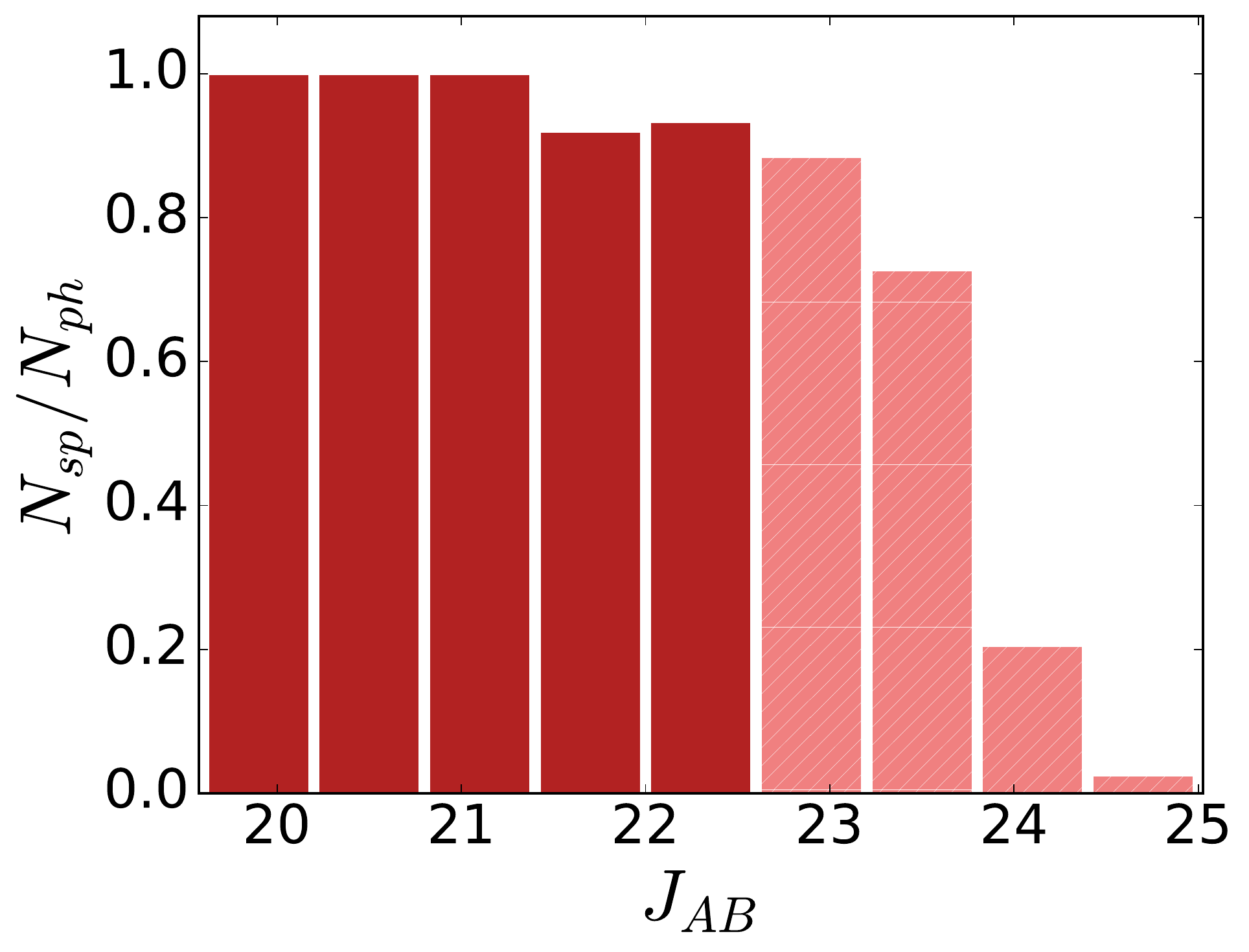}
\figcaption {G102 completeness, given as the ratio of spectroscopic to photometric detections, as a function of 3D-HST $J_{AB}$ magnitude. Solid bars indicate galaxies above our $90\%$ spectroscopic completeness limit, which occurs at $J_{AB}=22.6$.}
\label{Fig:sp_completeness}
\end{figure}

We estimated the spectroscopic and mass completeness of our sample by comparing the G102 sample with the much deeper photometric catalog in the spatial region of the protocluster. We find that 90\% of the photometric sources with $J_{AB}\leq22.6$ have extracted grism spectra (Figure~\ref{Fig:sp_completeness}). This limiting magnitude was then used to obtain an empirical estimate of the spectroscopic mass completeness, following the procedure of \citealt{Marchesini09}. To do this, we selected \textit{photometric} sources with $22.0\leq J_{AB} \leq 23.0$ and redshifts $1.40\leq z \leq 1.75$, and scaled their $J_{AB}$ luminosities to the $J_{AB}\leq22.6$ spectroscopic limit, holding galaxy mass-to-light ratio ($M/L$) constant.  Under the assumption that the distribution of galaxy $M/L$ is unchanged over the magnitude and redshift ranges we used, the luminosity scaling process creates a simulated, relatively complete population at the spectroscopic limit. The high-mass end of the simulated population then gives an estimate of the mass completeness. We find our $95\%$ grism mass-completeness limit to be $10^{10.2} M_{\odot}$, and we note that moderate changes to this limit ($10^{10.2\pm 0.1} M_{\odot}$) do not impact the results of this study. In the protocluster, $14/25$ extracted grism sources are above the mass completeness limit, and in the field $12/38$ sources are above the cutoff.

\section{Results}
\label{Sec:results}
\subsection{The Quiescent Fraction in IRC 0218 and the Field}
\label{Sec:uvj}
\begin{figure*}[htb!]
\begin{raggedleft}
\epsscale{1.1}
\plottwo{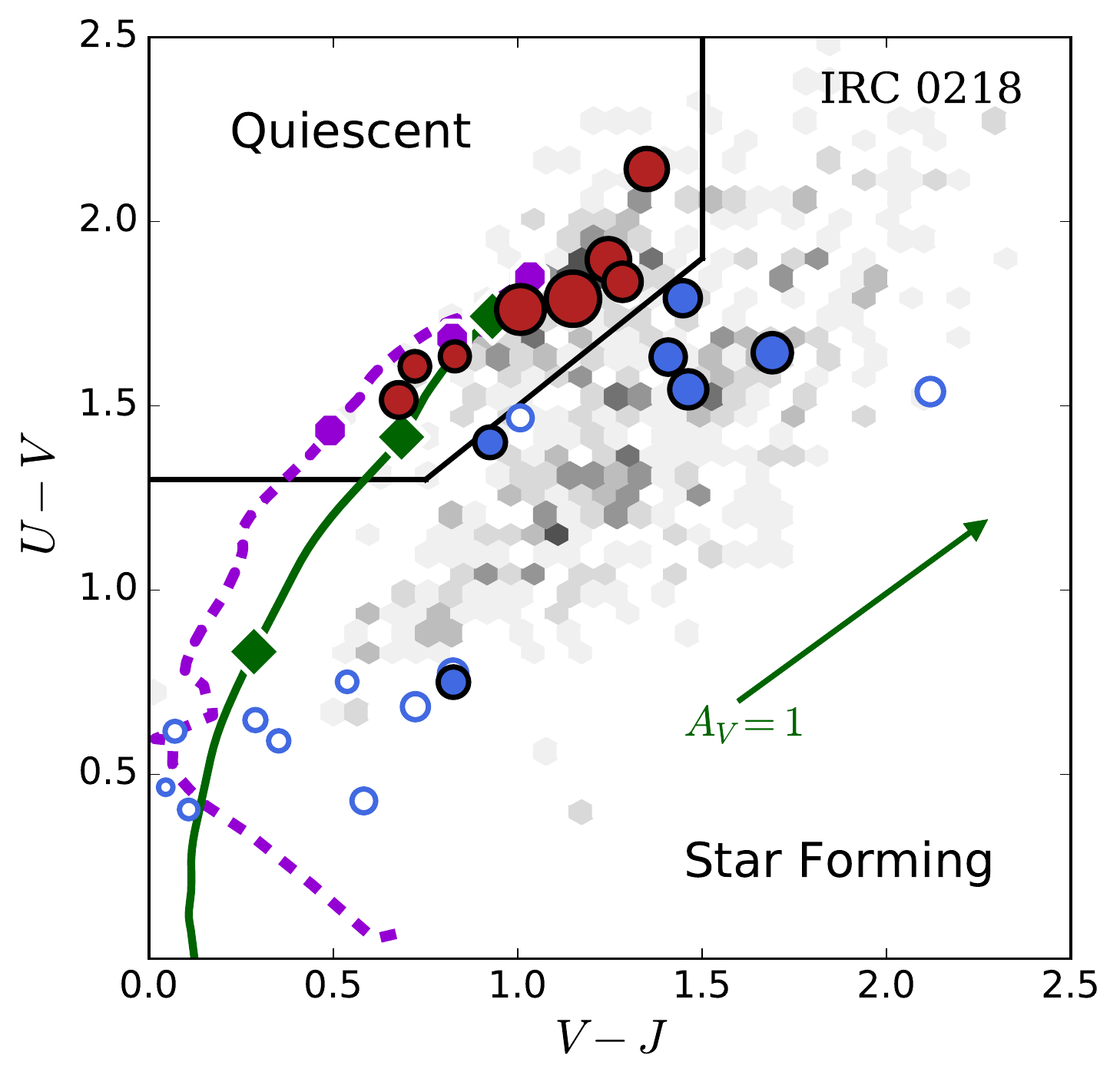}{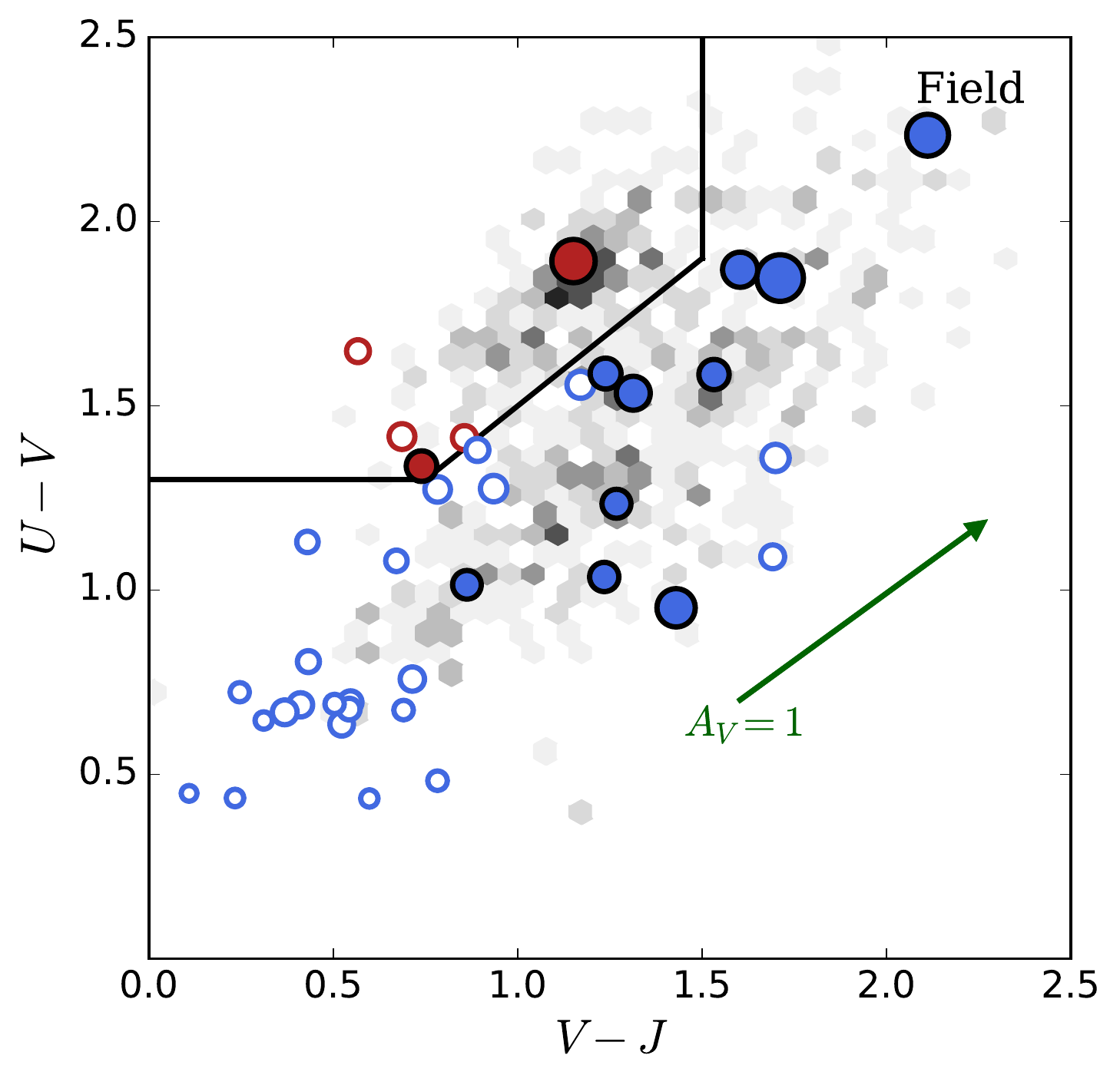}
\end{raggedleft}
\figcaption {$(U-V)$, $(V-J)$ color-color diagrams for the protocluster and field samples. Points in the upper left bounded region of each plot are classified as quiescent, while points outside the region are classified as star forming. Point size indicates mass logarithmically scaled relative to the most massive quiescent protocluster galaxy, $M_* =10^{11.48} M_{\odot}$. Open points represent galaxies below the adopted mass completeness threshold of $M_* = 10^{10.2} M_{\odot}$. The grayscale region shows the colors of the full 3D-HST G141 grism sample above the mass completeness threshold in the redshift range $1 \leq z \leq 2$. In the left panel, the dashed line traces the rest-frame color evolution for a $Z=0.02$ single stellar population from 10 Myr to 2.2 Gyr, as predicted by \citet{Bruzual03} SED models. The solid line traces evolution over the same period of time for a population with an exponential SFH with $\tau = 300$ Myr. Successive points on the evolution tracks (from left to right) indicate the predicted $z=1.62$ colors for populations with $z_f=$ 2.0, 2.5, and 3.5. The arrows indicate the change in $(U-V)$ and $(V-J)$ expected for a one magnitude increase in reddening in $V$, following a \citet{Calzetti00} extinction law.} 
\label{Fig:uvj_diagrams}
\end{figure*}

The protocluster and field samples were separated into quiescent and star forming populations according to their positions in the $(U-V)$,$(V-J)$ rest-frame color plane. Separating galaxies in the $UVJ$ plane allows for a distinction between quiescent galaxies and dust-obscured, star forming galaxies that might appear quiescent based on their $(U-V)$ colors alone (see \citet{Wuyts07, Williams09}). Here, we follow the criteria established by \citet{Whitaker12} and used in \citet{Whitaker13}, and define galaxies to be quiescent if they are contained in the region $(U-V) > 0.8 \times (V-J) + 0.7$, $(U-V) > 1.3$, $(V-J) < 1.5$. The results of this selection process are given in Figure~\ref{Fig:uvj_diagrams}. Several galaxies in the field and cluster are quite close to the selection boundary, and we discuss the sensitivity of our results to the specific adopted $UVJ$ cuts at the end of this section. 

We find that above our mass completeness limit, $8/14$ protocluster and $2/12$ field galaxies meet our adopted requirements for quiescence. Defining the quiescent fraction as $f_Q=n_{Q}/(n_{SF}+n_{Q})$, where $n_{Q}$ and $n_{SF}$ are the numbers of quiescent and star forming galaxies, respectively, we find that $f_Q = 0.57^{+0.15}_{-0.17}$ in the protocluster, and $f_Q = 0.17^{+0.18}_{-0.11}$ in the field. The upper and lower uncertainties given are the $68\%$ binomial confidence intervals calculated according to \citet{Gehrels86}. 

As our definition of quiescence only uses $UVJ$ information and our stellar mass values are derived from photometry, we extended the field sample to include all sources from the five 3HDST fields (UDS, GOODS-N, GOODS-S, COSMOS, and AEGIS), in order to avoid the larger uncertainties due to the small size of the G102 field sample. This extended field sample has 1151 galaxies with masses $M_{*}\geq 10^{10.2} M_{\odot}$ over the redshift range $1.52 \leq z \leq 1.76$, after exclusion of known IRC 0218 galaxies. Of these galaxies, 366 are quiescent and 785 are star forming, according to their $UVJ$ plane positions. From this sample we obtain $f_Q = 0.32^{+0.01}_{-0.01}$, consistent with the G102 field sample, though considerably more precise. Thus, we conclude that considering all galaxies with $M_* \ge 10^{10.2} M_{\odot}$, the protocluster is approximately $2 \times$ as quenched as the field. 

\begin{deluxetable}{ccccc}
\tablecaption{Mass, $f_{Q}$, and $f_{C}$}
\tablenum{1}
\label{Tab:clusteruvj}
\tablehead{\colhead{$log(M/M_\odot)$} & \colhead{$N_{Q}$} & \colhead{$N_{SF}$} & \colhead{$f_Q$} & \colhead{$f_C$}}
\startdata
\multicolumn{5}{c}{\bf{IRC 0218}}\\
11.175 & 4 & 0 & $1.00^{+0.00}_{-0.37}$ & $1.00^{+0.00}_{-0.63}$ \\
10.525 & 4 & 6 & $0.40^{+0.20}_{-0.18}$ & $0.33^{+0.23}_{-0.23}$ \\
\hline
\multicolumn{5}{c}{\bf{Field} \normalfont{(coarse mass bins)}}\\
11.175 & 108 & 132 & $0.45^{+0.03}_{-0.03}$ & \nodata \\
10.525 & 258 & 653 & $0.28^{+0.02}_{-0.02}$ & \nodata \\
\hline
\multicolumn{5}{c}{\bf{Field} \normalfont{(fine mass bins)}}\\
11.3 & 23 & 30 & $0.43^{+0.08}_{-0.08}$ & \nodata \\ 
11.1 & 35 & 47 & $0.43^{+0.06}_{-0.06}$ & \nodata \\ 
10.9 & 67 & 78 & $0.46^{+0.04}_{-0.04}$ & \nodata \\ 
10.7 & 101 & 153 & $0.40^{+0.03}_{-0.03}$ & \nodata \\ 
10.5 & 84 & 206 & $0.29^{+0.03}_{-0.03}$ & \nodata \\ 
10.3 & 56 & 271 & $0.17^{+0.02}_{-0.02}$ & \nodata \\ 
\enddata
\tablecomments{Only the mass-complete protocluster sample considered here. The field sample is drawn from all five 3D-HST fields. Masses correspond to bin centers; bin widths are $\Delta log(M/M_\odot) = 0.65$ for the protocluster and coarsely-binned field sample and $\Delta log(M/M_\odot) = 0.20$ for the field sample with finer mass binning. In order for the fine bins to encompass the same mass range as the coarse binning scheme, the bin at $log(M/M_\odot) = 11.3$ extends to $log(M/M_\odot) = 11.5$. The conversion fraction $f_C$ is calculated relative to the larger 3D-HST field sample.}
\end{deluxetable}

\begin{figure}[tb!]
\epsscale{1.17}
\plotone{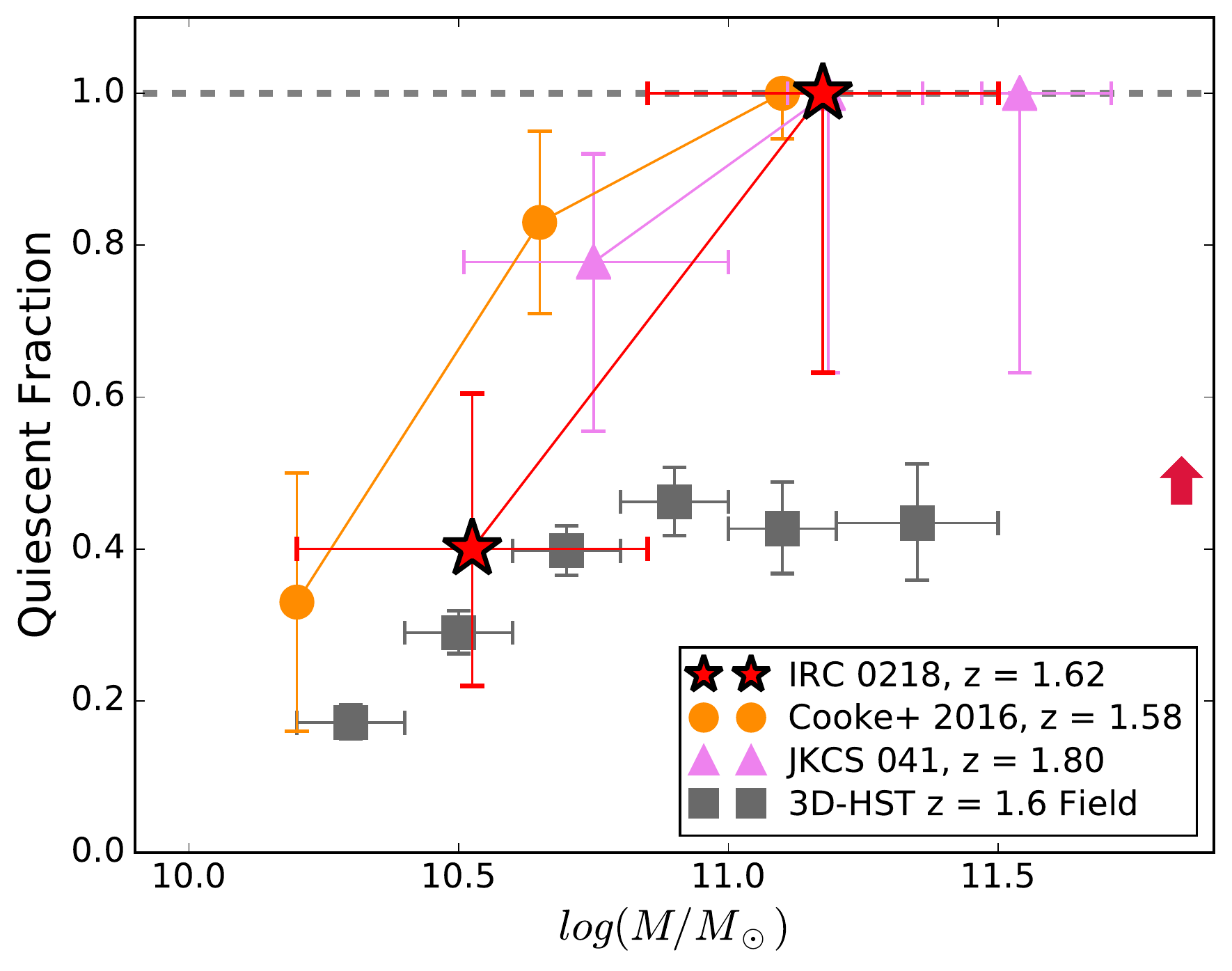}
\figcaption {The quiescent fraction of galaxies as a function of stellar mass from this work (red stars). We also give the quiescent fractions from the \citet{Newman14} study of the $z=1.80$ cluster JKCS 041 (pink triangles) and the work by \citet{Cooke16} on a $z=1.58$ cluster near the radio galaxy 7C 1753+6311 (orange circles) (see Section~\ref{Sec:discussion}). The field sample (gray squares) is composed of all galaxies in the full 3D-HST G141 grism sample with redshifts $1.52 \leq z \leq 1.76$, excluding known IRC 0218 galaxies. The arrow on the right side of the plot indicates the approximate change in $f_Q$ if quiescent galaxies were selected according to the $UVJ$ criteria of \citet{Williams09}.}
\label{Fig:quiescent_fraction_overlay}
\end{figure}

More interesting is examining how $f_Q$ varies as a function of stellar mass for the field and protocluster. For the protocluster sample, we calculated mean $f_Q$ in two mass bins, where the boundary between the bins was taken to be the midpoint of the range in mass spanned by the most massive protocluster galaxy and the mass completeness limit, $M_{split} = 10^{10.85} M_{\odot}$. For the field, galaxies were separated into 6 mass bins spanning $10^{10.2} \leq M_{*} \leq 10^{11.5}$. The results of this process are given in Table~\ref{Tab:clusteruvj}, and shown in Figure~\ref{Fig:quiescent_fraction_overlay}. We also calculated the cluster conversion fraction, or efficiency of quenching due to environment, according to the relation $f_{C} =(f_{Q, cluster} - f_{Q, field}) / (1 - f_{Q, field})$ \citep{vandenBosch08, Phillips14, Balogh16, Nantais17}. We give these values in Table~\ref{Tab:clusteruvj} to facilitate comparison with other cluster studies, but leave interpretation of our values for future work. 

We find evidence suggesting that the values of $f_Q$ in the cluster and field diverge at high masses. At low masses, $10^{10.2} M_{\odot} \le M_* \leq 10^{10.85} M_{\odot}$, we find $f_Q=0.40^{+0.18}_{-0.20}$ for the protocluster sample. This value is consistent with the field sample over the same mass range, where we find $f_Q=0.28^{+0.02}_{-0.02}$. At higher masses, however, $10^{10.85} M_{\odot} \leq M_* \la 10^{11.5} M_{\odot}$, $f_Q=1.00^{+0.00}_{-0.37}$ in the protocluster; the field value is $f_Q=0.45^{+0.03}_{-0.03}$. Thus, under the specific adopted binning scheme, the protocluster $f_Q$ is marginally consistent with enhancement above the field value at high masses. 

With regards to the stability of this result, we are hampered by the small protocluster galaxy sample size and correspondingly large binomial uncertainties. In both the high and low mass cluster bins, the median galaxy mass is similar (within $\sim 0.1$ dex) to the value for the respective field bin, so we do not believe our results are biased by differences in galaxy mass function between the protocluster and field. However, our results are relatively sensitive to the adopted binning scheme used for calculating $f_Q$. For example, if we adopt bins that encompass equal numbers of protocluster galaxies, rather than bins encompassing equal ranges in mass, then at high masses the protocluster and field $f_Q$ values are within $1 \sigma$ of one another. Thus, we avoid placing strong emphasis on the $f_Q$ results from our protocluster alone, and instead interpret our results in light of other high-redshift dense environments in Section~\ref{Sec:discussion}.

We also note that the exact value of $f_Q$ depends on the $UVJ$ criteria that are adopted. For example, if we adopt the slightly different \citet{Williams09} $UVJ$ criteria, three star forming galaxies above our mass limit - two in the protocluster and one in the field - would be reclassified as quiescent. Under such a change, we would find  $f_Q = 0.60^{+0.18}_{-0.21}$  and $f_Q = 0.32^{+0.02}_{-0.02}$ for the low mass protocluster and field bins, respectively. At high masses, under the \citet{Williams09} criteria, we would find $f_Q = 1.00^{+0.00}_{-0.37}$ and $f_Q = 0.52^{+0.03}_{-0.03}$ in the protocluster and field, respectively. Therefore, our conclusions regarding the behavior of $f_Q$ at high masses are stable against moderate changes in $UVJ$ criteria, but at lower masses some caution must be exercised, due to possible $UVJ$ selection effects.

\subsection{$D_{n}(4000)$ and Stellar Mass}
\label{Sec:d4km}
\begin{figure*}[htb!]
\begin{raggedleft}
\epsscale{1.1}
\plottwo{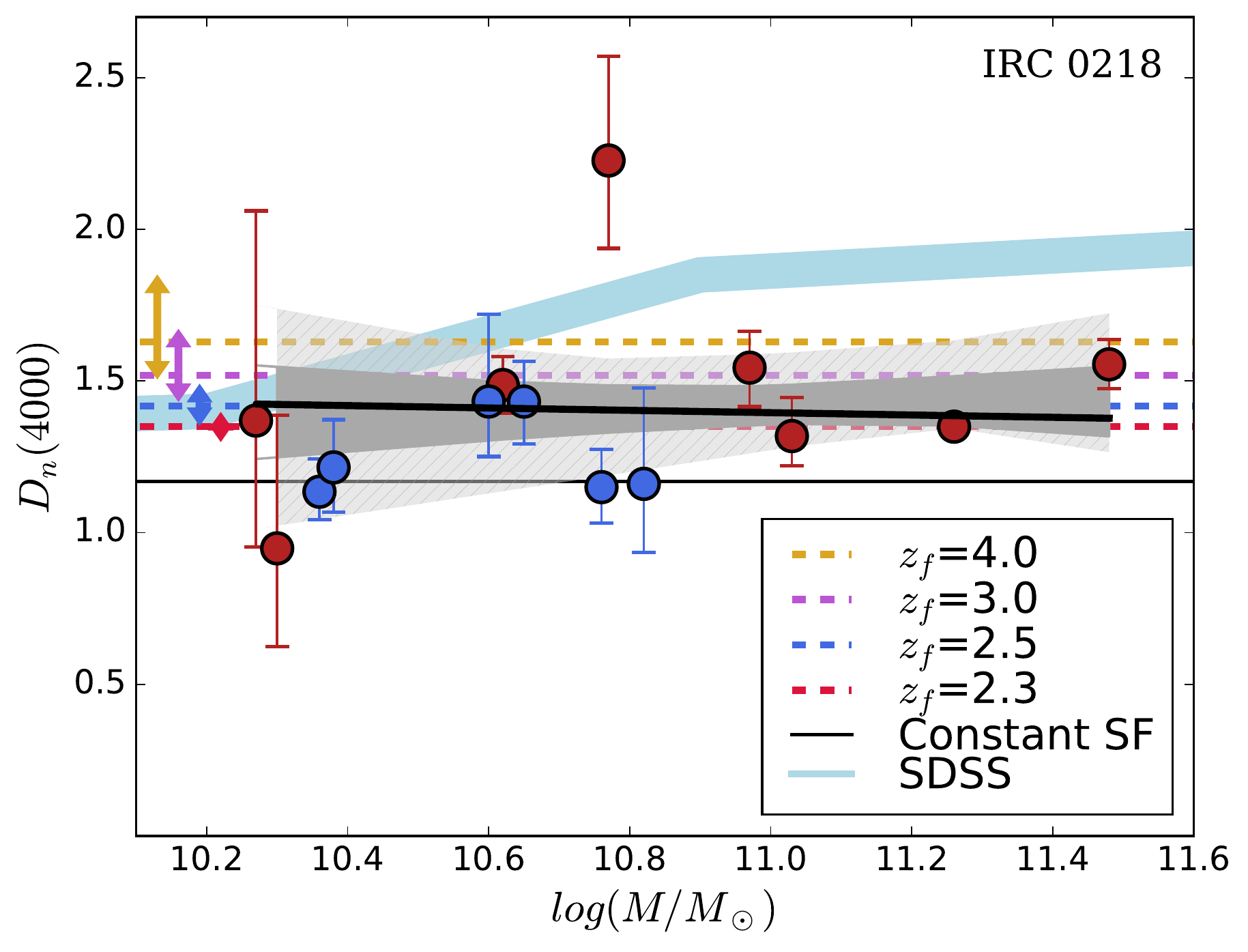}{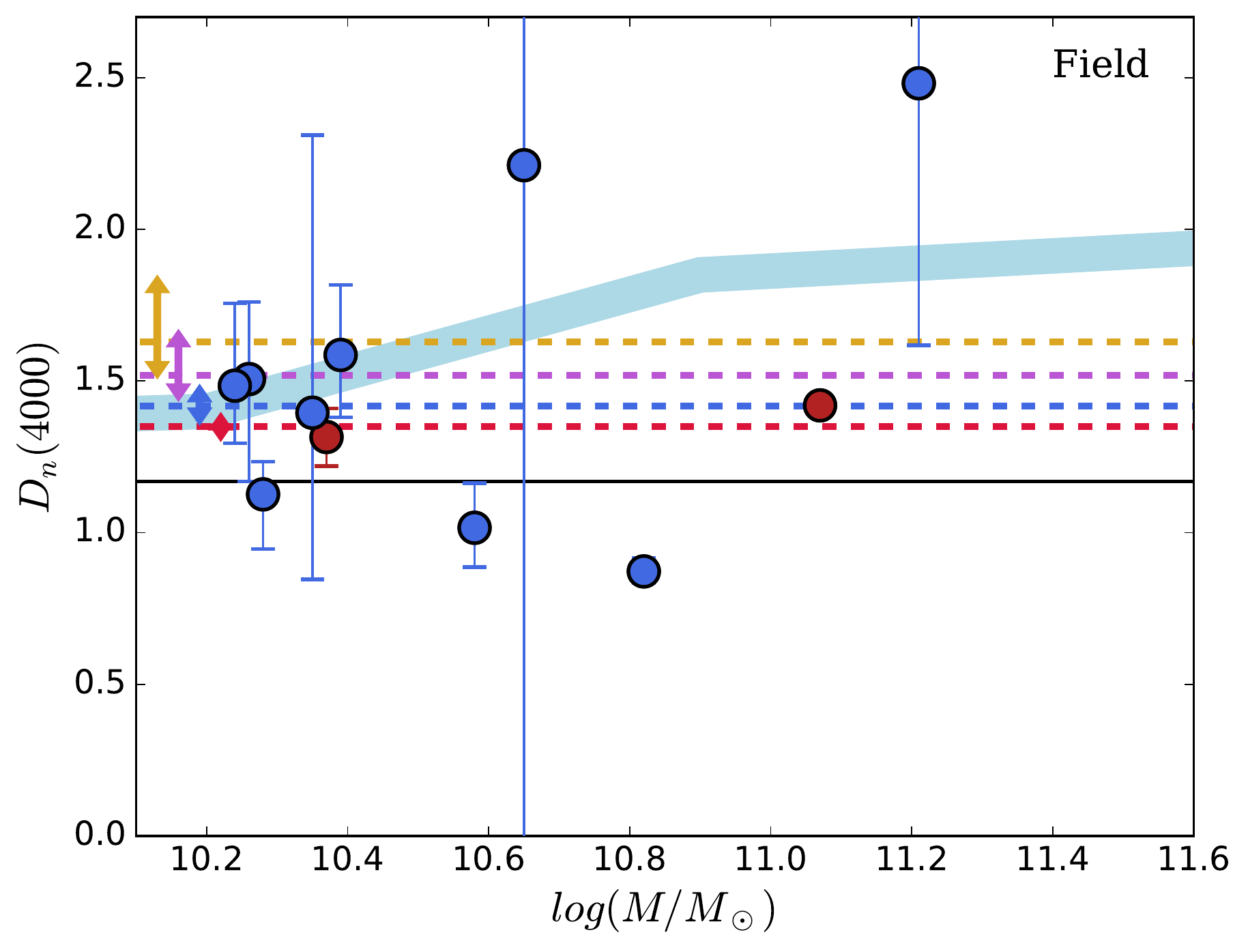}
\end{raggedleft}
\figcaption {$D_{n}(4000)$ plotted against stellar mass for the protocluster (left) and field (right) G102 grism samples. Red and blue points represent quiescent and star forming galaxies, respectively, as determined by galaxy $UVJ$ colors (see Figure~\ref{Fig:uvj_diagrams}). The thick blue line shows the approximate mean $D_{n}(4000)$ from an analysis of SDSS data by \citet{Kauffmann03}. The successive horizontal lines show $D_{n}(4000)$ predicted by \citet{Bruzual03} galaxy spectral evolution models with $Z=0.02$ and exponentially declining star formation histories ($\tau=300$ Myr), for a variety of formation redshifts. Arrows attached to the exponential SFH lines indicate the change in $D_{n}(4000)$ produced by changing galaxy metallicity to $Z=0.05$ or $Z=0.004$. In the left-hand panel, we show a weighted linear regression fit to $D_{n}(4000)$ as a function of mass for the quiescent protocluster galaxies, as described in Section~\ref{Sec:d4km}. The solid (hatched) shaded region around the fit shows the upper and lower $68\%$ ($95\%$) confidence interval on the fit. \textit{Note:} one star-forming field galaxy from Table~\ref{Table:field} (ID: 32468) has a negative $D_{n}(4000)$ due to poorly constrained redshift and is omitted from the figure.}
\label{Fig:d4kmass_diagrams}
\end{figure*}

We now turn our attention to $D_{n}(4000)$ and its ability to constrain age differences between galaxies in IRC 0218. In Figure~\ref{Fig:d4kmass_diagrams} we show $D_{n}(4000)$ as a function of stellar mass for the protocluster and field G102+G141 samples. The process by which $D_{n}(4000)$ was measured is detailed in Section~\ref{Sec:d4kmeasure}. Here, our mass-complete field sample consists of the 12 galaxies in the spatial region covered by the G102 grism, with redshifts $1.27\leq z \leq 1.58$, $1.66 \leq z \leq 1.77$. The cluster sample is the same as in Section~\ref{Sec:uvj}.

We first compare the mean $D_{n}(4000)$ values for quiescent and star forming protocluster galaxies with mean values for field galaxies. We find that the weighted mean $D_{n}(4000)$ for quiescent galaxies in the protocluster is ${\mu}_Q = 1.39 \pm 0.03$, where the weight associated with each $D_{n}(4000)$ value is taken as the estimate of inverse variance derived from the measure's $68\%$ confidence interval. For the star forming galaxies in the cluster, ${\mu}_{SF} = 1.22 \pm 0.06$. In the field, we find that ${\mu}_Q = 1.41 \pm 0.03$; this is consistent with the cluster value, though we note that our field sample only contains 2 quiescent galaxies. However, ${\mu}_{SF} = 0.94 \pm 0.04$ in the field, driven by a galaxy with a very low, well-constrained $D_{n}(4000)$ value. Since we report weighted mean $D_{n}(4000)$ values here, this measurement dominates the reported ${\mu}_{SF}$. Removing it from the calculation results in ${\mu}_{SF} = 1.20 \pm 0.08$ in the field. 

Such a dramatic change indicates that more field galaxies with accurate $D_{n}(4000)$ measurements are needed before drawing any conclusions regarding any differences between the star forming field and protocluster populations. Likewise, differences in the scatter of $D_{n}(4000)$ measurements in the different environments could potentially yield important information as to the relative SFHs of cluster and field galaxies. However, the small numbers of galaxies in our samples preclude the precise measurement of the scatter. Future larger samples of $D_{n}(4000)$ will be needed to do this.

Using the same mass bins as in Section~\ref{Sec:uvj}, and now considering all protocluster galaxies (both quiescent and star forming), we find the weighted mean $D_{n}(4000)$  at stellar masses $10^{10.85} M_{\odot} \leq M_* \la 10^{11.5} M_{\odot}$ to be ${\mu}_{high} = 1.38 \pm 0.03$. For masses $10^{10.2} M_{\odot} \leq M_* \leq 10^{10.85} M_{\odot}$, we find ${\mu}_{low} = 1.31 \pm 0.05$. Thus, there is little difference in mean $D_{n}(4000)$ between the two mass ranges. This result is unchanged under changes to the mass binning scheme. For example, adopting the equal-number binning discussed in Section~\ref{Sec:uvj} would result in closer agreement between the high and low mass subsamples. 

The $D_{n}(4000)$ values we calculate for IRC 0218 galaxies are in general much lower than those observed in the local Universe. In Figure~\ref{Fig:d4kmass_diagrams}, we show the approximate locus of $D_{n}(4000)$ values from an analysis of Sloan Digital Sky Survey (SDSS) galaxies by \citet{Kauffmann03}. At high masses, the mean $D_{n}(4000)$ value in the protocluster corresponds to the lowest few percent of $D_{n}(4000)$ values for the entire SDSS sample. At lower masses, the mean protocluster value is still lower than for the SDSS galaxies. These results indicate that as expected, the galaxies we observe at $z=1.6$ contain much younger stellar populations than at $z\sim 0$. We discuss the implications of these results under certain assumptions in Section~\ref{Sec:discussion}.

A primary goal of this study was to place constraints on any relation between red sequence galaxy age and mass; we therefore used weighted linear regression to fit the red sequence $D_{n}(4000)$ values as a function of stellar mass for protocluster galaxies. Weights were again chosen to be the estimate of inverse variance derived from the average $68\%$ confidence interval for each galaxy. Uncertainties on the fit were determined by bootstrapping. The results of our regression analysis are given in Figure~\ref{Fig:d4kmass_diagrams}. Within our $68\%$ uncertainties associated with our fit, we are consistent with slopes ranging from $\Delta D_{n}(4000)=+0.25$ log$(M_{\odot})^{-1}$ to $\Delta D_{n}(4000)=-0.2$ log$(M_{\odot})^{-1}$ over our mass interval. 

Thus, we cannot reject the null hypothesis that there is no relation between $D_{n}(4000)$ and stellar mass for quiescent galaxies above our mass completeness limit. Depending on adopted star formation history, $D_{n}(4000)$ can begin to saturate as early as $D_{n}(4000) \sim 1.6$ for solar-metallicity models, losing its ability to differentiate between galaxies of different age. This saturation threshold is somewhat lower for lower metallicities. However, few of the protocluster galaxies have $D_{n}(4000)$ values close to this level; we can therefore rule out saturation as the underlying cause for a flat relation, unless the galaxies we observe have $Z<Z_{\odot} /5$, which is unlikely given the relatively high stellar mass range posed here and based on the metallicity of star-forming galaxies in this same protocluster \citep{Tran15}. The implications of a flat relation for the stellar age distribution of the protocluster galaxies under certain model assumptions are discussed further in Section~\ref{Sec:discussion}. 

Our $D_{n}(4000)$ results are robust against moderate changes to the underlying $UVJ$ quiescent/star forming classifications. Under the reclassification of the three galaxies discussed at the end of Section~\ref{Sec:uvj}, the significance of the difference between star forming and quiescent galaxies in IRC 0218 would be essentially unchanged: $\mu_{Q} = 1.39 \pm 0.03$ and $\mu_{SF} = 1.17 \pm 0.07$. Additionally, under such a reclassification, we still cannot reject the null hypothesis of no relation between $D_{n}(4000)$ and galaxy mass for quiescent protocluster galaxies. 

\section{Discussion}
\label{Sec:discussion}
\subsection{The Quiescent Fraction in IRC 0218}
As discussed in Section~\ref{Sec:uvj} and shown in Figure~\ref{Fig:quiescent_fraction_overlay}, in IRC 0218, the quiescent fraction at $M_{*}\geq 10^{10.85} M_{\odot}$ is consistent with being 2$\times$ higher than in the field. While the strength of this conclusion is hampered by a small sample size, it is similar to the results from clusters at similar mass and redshift \citep{Newman14, Cooke16}. Considering these three dense environments together, the implication is that even at the high redshifts of these clusters, the environment has already played some role in populating the red sequence. 

Comparing these results with those of \citet{Nantais16}, which we do not show in Figure~\ref{Fig:quiescent_fraction_overlay} due to differences in $UVJ$ selection criteria, we find that the quiescent fractions plotted in Figure~\ref{Fig:quiescent_fraction_overlay} are systematically higher at high masses by $\Delta f_{Q} \sim 0.2 - 0.3$. It is not immediately clear what may be driving this discrepancy. As noted, the $UVJ$ selection criteria for quiescence are slightly different in \citet{Nantais16}, but adopting their criteria actually worsens the discrepancy - the cluster quiescent fractions remain relatively unchanged since most galaxies are well-separated from the selection boundary, but the field fraction actually decreases on average.

Alternately, the discrepancy may just be a reflection of large cluster-to-cluster variance at high redshift. Both \citet{Nantais16} and \citet{Hatch16} note that environmental quenching efficiency may depend on the time galaxies have spent in the cluster environment, and several studies have found that star formation activity varies appreciably even among similarly-selected clusters \citet{Brodwin13, Alberts16}. Since the main group in IRC 0218 (which we examine in this work) appears to be more evolved, on average, than similar-redshift protoclusters \citep{Hatch16}, and JKCS 041 has a very well-developed red sequence for its redshift \citet{Newman14}, the discrepancies in the high-mass quiescent fraction may be driven by the specific assembly histories of the clusters being compared. That is, the high-redshift environments we show in Figure~\ref{Fig:quiescent_fraction_overlay} may be more evolved than the \citet{Nantais16} clusters. Indeed, it may be that clusters selected by different methods may have systematically different quenched fractions.

At masses below $M_{*}=10^{10.85} M_{\odot}$, we find that the quiescent fraction in IRC 0218 is consistent with the field value at $z \sim 1.6$. This mirrors the trends observed in the clusters examined by \citet{Newman14} and \citet{Cooke16}, and is also consistent with the results of \citet{Nantais16}. It therefore appears that if there is an environmental effect on the assembly of the red sequence at these redshifts, it is mass-dependent, primarily impacting high-mass galaxies with $M_{*}\geq 10^{10.85} M_{\odot}$, at least at $z > 1.5$. 

\begin{figure}[htb!]
\epsscale{1.17}
\plotone{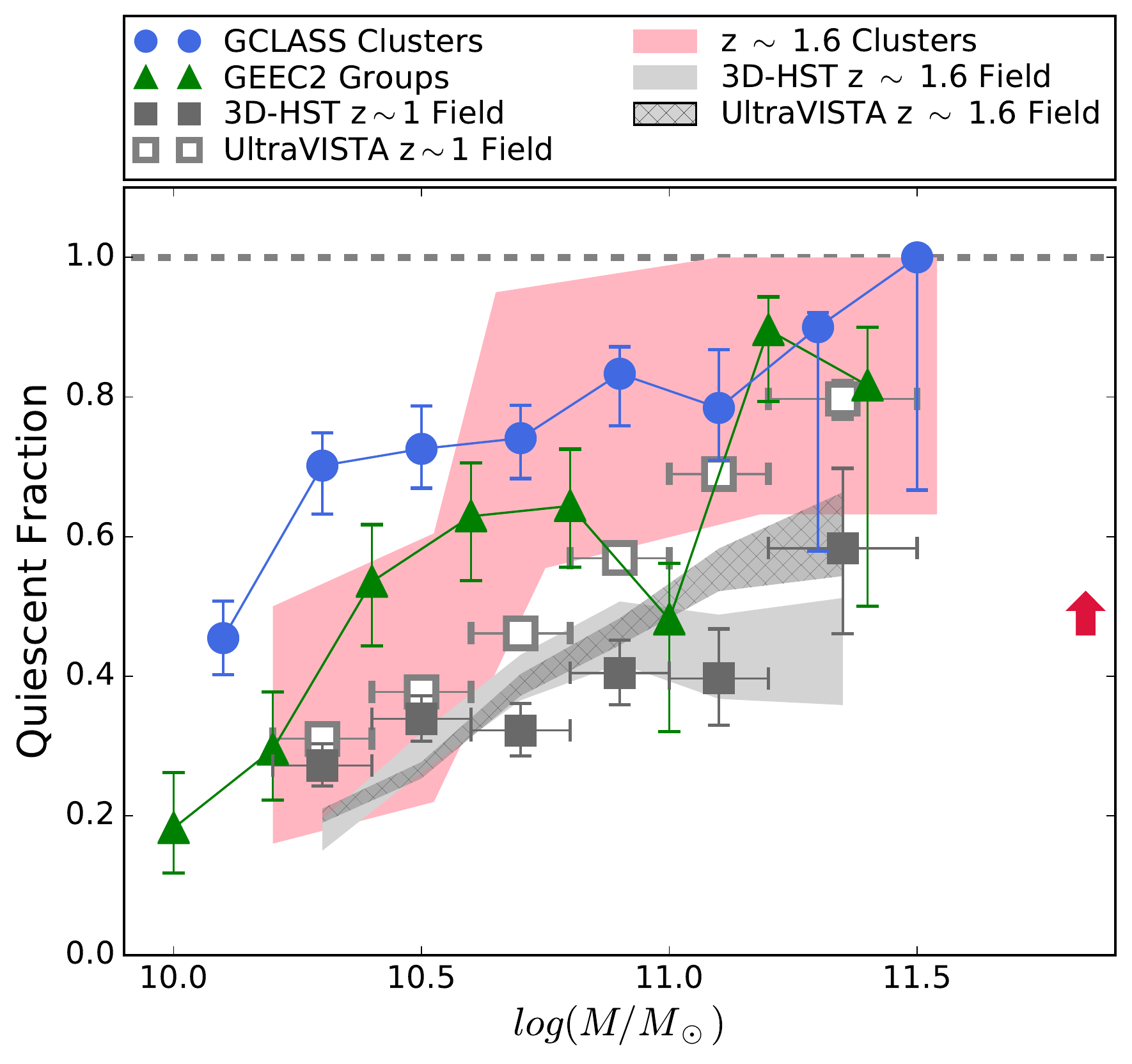}
\figcaption {The quiescent fraction of galaxies as a function of stellar mass from the GEEC2 group and GCLASS cluster samples \citep{Balogh16} at $z\sim 1$. The field sample (solid gray squares) here consists of 3D-HST galaxies with masses $M_{*}
\ge 10^{10.2} M_{\odot}$ and redshifts $0.9\leq z \leq 1.1$. The solid background regions correspond to the envelopes traced out by the 68\% binomial confidence limits for the higher-redshift dense environment (light red) and field (gray) samples from Figure~\ref{Fig:quiescent_fraction_overlay}. The hatched region corresponds to $f_Q$ for UltraVISTA galaxies chosen using the $UVJ$ criteria given in \citet{Muzzin13}. The open points show $f_Q$ for UltraVISTA galaxies chosen in the same way as the 3D-HST $z\sim 1$ field sample. We note that for the field samples, the stellar masses plotted were chosen to be consistent with the $M_{*} = 10^{10.2} M_{\odot}$ IRC 0218 stellar mass limit. The red arrow indicates the approximate change in values of $f_Q$ derived from 3D-HST galaxies if the $UVJ$ criteria of \citet{Williams09} were adopted.}
\label{Fig:quiescent_fraction_overlay2}
\end{figure}

To look for evidence of an evolution in $f_Q$ at high redshifts, we compared our quiescent fraction results with those from the GEEC2 group and GCLASS cluster surveys at $z\sim 1$ \citep{Balogh16}. We find no evidence that the  quiescent fraction in dense environments at stellar masses ($M_{*}\geq 10^{10.85} M_{\odot}$) evolves as a function of redshift over 1 $\leq z \leq 1.6$. as seen in Figure~\ref{Fig:quiescent_fraction_overlay2}, at least for the clusters shown. This is at odds with the work by \citet{Nantais17}, where environmental quenching efficiency was shown to strongly evolve between $z=1.6$ and $z=0.9$. However, the apparent absence of evolution at high masses that we find would be consistent with the scenario discussed previously, in which the three high-redshift clusters given in Figure~\ref{Fig:quiescent_fraction_overlay} were more evolved than the average $z\sim 1.6$ cluster (or conversely, that the SpARCS clusters \citep{Nantais17} were less-evolved than average). Observations of more high-$z$ clusters are needed to better understand the high variation in cluster-to-cluster properties, and may help to resolve the relative evolutionary state of IRC 0218.

Drawing conclusions at lower halo masses is difficult, due to the uncertain masses of the high redshift clusters and small number statistics, but IRC 0218 is likely a progenitor of a typical $M_{*}=10^{14} - 10^{14.5} M_{\odot}$ cluster at $z=1$ \citep{Rudnick12}, i.e., its likely descendent would be a GCLASS cluster, rather than a GEEC2 group. If this is so, then the lower stellar mass end would appear to evolve in its quiescent fraction over 1 $\leq z \leq 1.8$, increasing by a factor of $\sim 1.5 \times$ for galaxies of mass $10^{10.2} M_{\odot} \leq M_* \leq 10^{10.85} M_{\odot}$. The JKCS 041 sample does not extend to low enough masses to test this conclusion, but the $z=1.5$ \citet{Cooke16} cluster is consistent with our inference from IRC 0218. 

Interpreting the behavior in the field over the same redshift range is more difficult. If we compare the 3D-HST $z \sim 1.6$ field sample detailed in Section~\ref{Sec:uvj} to a sample consisting of 3D-HST galaxies with redshifts $0.9 \la z \la 1.1$, we see no evidence for evolution in $f_Q$ over the redshift range $1 \la z \la 1.6$ (see Figure~\ref{Fig:quiescent_fraction_overlay2}). On the other hand, we can instead construct the $z \sim 1.6$ and $z \sim 1$ field samples using UltraVISTA galaxies from the version 4.1 catalog release \citep{Muzzin13}. For the UltraVISTA field sample, we adopt the slightly different, redshift-dependent $UVJ$ criteria described in \citet{Muzzin13}, in order to better separate the UltraVISTA red sequence galaxies (see below). At $z\sim 1.6$ we find that the UltraVISTA field is consistent with the 3D-HST field sample over our mass range, but at $z\sim 1$, the UltraVISTA $f_Q$ values are systematically higher over all masses. 

The largest contributor to the $z\sim 1$ discrepancy between 3D-HST and UltraVISTA appears to be a relatively large offset in rest-frame color between the two surveys. This offset between the two surveys is only apparent at $z=1$, not at $z=1.6$, which indicates a redshift-dependent difference in the rest-frame colors of the two surveys. Use of the \citet{Muzzin13} $UVJ$ criteria to separate UltraVISTA galaxies lessens the differences in $f_Q$ at $z \sim 1$ somewhat relative to using the \citet{Whitaker12} criteria for both 3D-HST and UltraVISTA, but as shown in Figure~\ref{Fig:quiescent_fraction_overlay2}, the discrepancy is still significant, particularly at higher masses. We show the $UVJ$ diagrams for 3D-HST and UltraVISTA, the different $UVJ$ selection criteria, and the differences in rest-frame color between the two surveys in Appendix~\ref{Appendix2}, and note that this situation illustrates that extreme caution must be exercised when comparing $f_Q$ between different surveys.

Ultimately, we refrain from drawing any firm conclusions about $f_Q$ evolution in the field, due to the discrepancy between the 3D-HST and UltraVISTA field galaxy quiescent fractions. However, at low masses, both the 3D-HST and UltraVISTA field $z \sim 1$ $f_Q$ values are elevated relative to the $z=1.6$ values. If we were to take this mild quiescent fraction evolution in low stellar mass field galaxies at face value, it would contrast with the significant evolution seen in the clusters at low stellar masses over the same redshift range. This could imply that the cluster is more effective at quenching low mass galaxies than processes at play in the field \citep[e.g.,][]{Tomczak14}. More cluster studies and careful treatment of $UVJ$ selection effects are needed to evaluate this possibility.

\subsection{$D_{n}(4000)$ and Stellar Mass}

If the same star formation history (SFH) is assumed for all galaxies, a lack of $D_{n}(4000)$ dependence with stellar mass implies no age trend along the red sequence. An absence of such a trend could be explained by dry (gasless) merging, which would serve to redistribute stellar mass on the red sequence and thus scramble any trend in age, and thus $D_{n}(4000)$, with stellar mass. The evolutionary state of IRC 0218 implies a high merger rate, and evidence for an elevated rate has been found by previous studies \citep{Rudnick12, Papovich12, Lotz13}. \citet{Lotz13} found that most of the mergers were between passive galaxies, consistent with them being mostly dry (gasless) and therefore not hosting any new star formation. However, future work is needed to investigate the possible impacts of dry merging on red sequence $D_{n}(4000)$ through measurements of galaxy morphology and $D_{n}(4000)$ in more high-redshift clusters.

On the other hand, if merging is not important, then the mass dependence of $f_Q$ coupled with a lack of similar trend in $D_{n}(4000)$ indicates that the quenching of star formation may have happened at roughly the same epoch in such a way as to imprint the $f_Q - M_*$ trend. Our data are too limited to highly constrain such a scenario, but speculatively, such a quenching episode could be associated with the initial formation of the protocluster core, as our G102 data only span the densest region of IRC 0218. Clearly more $D_{n}(4000)$ measurements in high-redshift clusters are needed to distinguish the various processes playing a role in these environments. 

There is also the possibility that a true trend with stellar mass exists, as we are unable to rule out strong trends with slopes $\Delta D_{n}(4000)=+0.25$ log$(M_{\odot})^{-1}$ to $\Delta D_{n}(4000)=-0.2$ log$(M_{\odot})^{-1}$ over our mass range.  Indeed, there is a common expectation that more massive cluster galaxies should have, on average, older populations (e.g., \citealt{Rosati09, Jorgensen13, Tanaka13}). It may be that the relatively large uncertainties associated with our relation between $D_{n}(4000)$ and stellar mass (see Section~\ref{Sec:d4km}) mask an underlying trend or that $D_{n}(4000)$ is not sufficiently sensitive to age to reveal any trends in the IRC 0218 galaxies. 

We do note that the weighted mean $D_{n}(4000)$ for the red sequence galaxies in the protocluster is the same as the value found in our field sample. With the caveat that our field sample only contains 2 quiescent galaxies, this result implies that the field and cluster galaxies we examined did not quench at appreciably different times, though the cluster environment apparently quenched more efficiently at a given stellar mass, at least for masses  $M_* \geq 10^{10.85} M_{\odot}$. Simulations predict such ``accelerated evolution'' of cluster environments due to earlier collapse of dark matter halos \citep{deLucia04}. This is supported by more recent observations of both IRC 0218 \citep{Papovich12} and other environments \citep{Strazzullo13, Newman14}. Under such a scenario, even if more massive galaxies in dense environments quench earlier than lower mass galaxies, if evolution and/or quenching in dense environments is accelerated, then the age differences between galaxies may be too small to show statistically significant trend in $D_{n}(4000)$, given the uncertainties in our slope described previously. Demonstration of a strong $D_{n}(4000)$-$M_*$ relation in high-redshift clusters could place constraints on any potential accelerated evolution in dense environments.

Under assumptions regarding SFH, we can translate between $D_{n}(4000)$ and stellar age/formation redshift. In this work, we adopt a common, exponentially declining SFH with characteristic timescale $\tau=300$ Myr and metallicity $Z=0.02$. Using GALAXEV \citep{Bruzual03} to model galaxy spectral evolution under this SFH, we show the predicted $D_{n}(4000)$  values for various formation redshifts in Figure~\ref{Fig:d4kmass_diagrams}. We find that the quiescent galaxies in IRC 0218 are largely consistent with star formation starting in the redshift range $2.3 \la z_f \la 3$, consistent with the formation epoch estimated by \citet{Papovich10} on the basis of $(U-B)$ colors, and similar to the $z_f$ estimated by several other high-redshift cluster studies, as well as studies of $z < 1$ environments (e.g. \citealt{Bower98, Poggianti01, vanDokkum07, Mei09, Hilton09}).

Several galaxies are consistent with very high formation redshifts, $z_f \ga 4$; these high $z_f$ values are consistent with the $z\sim 1$ observations by \citet{Muzzin12}, who found quiescent galaxies in clusters with ages approaching the age of the Universe. Such high $z_f$ values place strong constraints on possible quenching mechanisms; any possible mechanism would need to be capable of truncating star formation in the limited time before $z\sim 3$. For example, a galaxy that formed stars until $z=4$ and was subsequently quenched by $z=3$ would have had the quenching occur over a period of only $\sim 600$ Myr. This limited time interval naturally disfavors known low-redshift quenching mechanisms operating on long timescales \citep[e.g., delayed quenching or strangulation,][]{Wetzel13, Peng15}, and favors more abrupt mechanisms \citep[e.g., luminous AGN feedback,][]{Bongiorno16}, in line with several other high-redshift cluster studies \citep{Brodwin13, Alberts16}.

Our results regarding $z_f$ are highly dependent on adopted SFH. If a common $\tau=300$ Myr, $Z=0.02$ SFH is appropriate then we can rule out red sequence formation ages more recent than $z=2.3$, as those galaxies would not meet our $UVJ$ cut, as shown in Figure~\ref{Fig:uvj_diagrams}. However, if instead we adopted a single stellar population (SSP) with the same metallicity, then we could not rule out formation redshifts as recent as $z\le2$. 

The effects of raising or lowering adopted galaxy metallicity are illustrated in in Figure~\ref{Fig:d4kmass_diagrams}. As can be seen, changing galaxy metallicity would impact our $z_f$ results; $D_{n}(4000)$ is relatively sensitive to metallicity, with higher metallicities leading to higher $D_{n}(4000)$ for a given stellar age. Full spectral fitting may alleviate some of the sensitivity to metallicity.

On the other hand, if metallicity were somehow masking a positive correlation between $D_{n}(4000)$ and stellar mass, then metallicity would need to be negatively correlated with stellar mass. Such an inverse relation is not observed in studies at both low and high redshifts (low redshift: \citealt{Tremonti04, Gallazzi05}, high redshift: \citealt{Erb06, Zahid14, Tran15}). Therefore, metallicity effects are likely not responsible for the flat $D_{n}(4000)$-$M_*$ relation we observe.

\section{Conclusion}
\label{Sec:conclusion}
We present the results of a study of 14 quiescent and star forming galaxies in the largest subgroup of the $z=1.62$ protocluster IRC 0218, down to stellar masses $M_{*}=10^{10.2} M_{\odot}$. Using IR photometry, \textit{Hubble}/WFC3 grism spectroscopy of the rest-frame $4000$ $\mathrm{\AA}$ break, and a mixture of spectroscopic and grism redshifts, we classified galaxies in the protocluster and field as star forming or quiescent based on their rest-frame $(U-V)$ and $(V-J)$ colors. Stellar masses were estimated through template fitting, and the spectroscopic index $D_{n}(4000)$ was measured for each galaxy.

Considering protocluster galaxies with $M_{*} \geq 10^{10.85} M_{\odot}$, we find the quiescent fraction in IRC 0218 to be $f_Q = 1.0^{+0.00}_{-0.37}$. This value is $2\times $ higher than the value in the field, which we find to be $f_Q = 0.45^{+0.03}_{-0.03}$, though there are only 4 galaxies in our protocluster sample with masses $M_{*} \geq 10^{10.85} M_{\odot}$. At lower masses, $10^{10.2} M_{\odot} \leq M_{*} \leq 10^{10.85} M_{\odot}$, the protocluster and field have consistent $f_Q$, with $f_Q = 0.40^{+0.20}_{-0.18}$ in the cluster and $f_Q = 0.28^{+0.02}_{-0.02}$ in the field. 

Though our conclusion of an elevated $f_Q$ at high masses in IRC 0218 is of marginal significance due to our small sample size, it is consistent with two other high-redshift cluster studies. However, we do note that there are other, similar-redshift clusters in the literature with high-mass quiescent fractions that are consistent with values in the field, though the clusters are all consistent within their formal uncertainties. It is therefore difficult to determine if the range in quiescent fractions reflects a true cluster-to-cluster variation or is dominated by statistical uncertainties.

Comparing our results and those from two other protoclusters with studies of groups and clusters at $z\sim 1$, we see no evolution in the quiescent fraction at masses $M_{*} \geq 10^{10.85} M_{\odot}$ between $z\sim 1.6$ and $z\sim 1$. On the other hand, there is one recently published sample of clusters at $z\sim 1.6$ that does show significant evolution to $z\sim 1$, again implying a large cluster-to-cluster variation in properties. For the low stellar mass end, $10^{10.2} M_{\odot} \le M_{*} \leq 10^{10.85} M_{\odot}$, comparing IRC 0218 with likely descendents at $z\sim 1 $ indicates that $f_Q$ may increase modestly between the two epochs. 

Our data are consistent with the null hypothesis of no trend between $D_{n}(4000)$ and stellar mass for quiescent protocluster galaxies above our mass limit. However, the data are also consistent at the $<2\sigma$ level with significant trends of $D_{n}(4000)$ with stellar mass. The null hypothesis is consistent with a scenario in which the high merger rate in IRC 0218 has redistrubuted red sequence mass, masking any previously existing trend in $D_{n}(4000)$. Alternately, it may suggest that the trend in $f_Q$ was imprinted as the protocluster red sequence galaxies quenched over a short period of time. Overall, we find that as expected, quiescent galaxies in the cluster and field have elevated $D_{n}(4000)$ relative to star forming galaxies. 

Under an adopted $\tau=300$ Myr, $Z=0.02$ exponentially declining SFH, we can rule out quiescent galaxies forming more recently than $z=2.3$. Most quiescent galaxies in the cluster formed in the redshift range $2.3 \la z_f \la 3$. Several red sequence galaxies have very high formation redshifts consistent with $z_f \ga 4$, placing strong timescale constraints on any potential quenching mechanism. 

This work is based on observations taken with the NASA/ESA $HST$ as part of the 3D-HST Treasury Program (GO 12177 and 12328), which is operated by the Association of Universities for Research in Astronomy, Inc., under NASA contract NAS5-26555. This work is also based on NASA/ESA $HST$ observations taken as part of program GO 12590, supported by NASA through a grant from the Space Telescope Science Institute. DLB gratefully acknowledges support from NSF grant AST-1211621. GR acknowledges the support of NASA grant HST-GO-12590.011-A, NSF grants 1211358 and 1517815, the support of an ESO visiting fellowship, and the hospitality of the Max Planck Institute for Extraterrestrial Physics, and Hamburg Observatory. K. Tran acknowledges support by the National Science Foundation under grant number 1410728. CNAW acknowledges the support of NASA grant HST-GO-12590.09-A. DLB and GR thank Katherine Whitaker and Adam Muzzin for helpful discussion. The authors thank the anonymous referee for constructive comments.

\bibliographystyle{aasjournal}
\bibliography{d4k_refs}

\appendix
\label{Appendix1}
\section{The IRC 0218 and Field Galaxy Grism Samples}

\begin{deluxetable*}{LcLccccR}[h!]
\tablehead{\colhead{3D-HST ID} & \colhead{Q/SF} & \colhead{$z$} & \colhead{$J_{AB_{rest}}$} & \colhead{$(U-V)_{rest}$} & \colhead{$(V-J)_{rest}$} & \colhead{$log(M/M_{\odot})$} & \colhead{$D_{n}(4000)$}}
\tablecaption{IRC 0218 Galaxy Sample}
\tablenum{2}
\startdata
31684 & Q & 1.631^{+0.003}_{-0.008} & 19.82 & 1.79 & 1.15 & 11.48 & 1.56^{+0.08}_{-0.08} \\ 
30737 & Q & 1.621 \tablenotemark{a,b} & 19.60 & 1.76 & 1.01 & 11.26 & 1.35^{+0.03}_{-0.03} \\ 
36010 & Q & 1.628^{+0.003}_{-0.003} & 20.39 & 1.90 & 1.25 & 11.03 & 1.32^{+0.13}_{-0.10} \\ 
29983 & Q & 1.629^{+0.003}_{-0.003} & 20.52 & 2.14 & 1.35 & 10.97 & 1.54^{+0.12}_{-0.13} \\ 
30169 & SF & 1.629 \tablenotemark{b} & 20.67 & 1.64 & 1.69 & 10.82 & 1.16^{+0.32}_{-0.23} \\ 
29899 & Q & 1.620^{+0.005}_{-0.006} & 20.98 & 1.84 & 1.28 & 10.77 & 2.23^{+0.34}_{-0.29} \\ 
30545 & SF & 1.624 \tablenotemark{a,b} & 20.71 & 1.54 & 1.46 & 10.76 & 1.15^{+0.12}_{-0.12} \\ 
29007 & SF & 1.624^{+0.004}_{-0.005} & 20.84 & 1.79 & 1.45 & 10.65 & 1.43^{+0.13}_{-0.14} \\ 
33092 & Q & 1.621^{+0.004}_{-0.004} & 21.19 & 1.52 & 0.68 & 10.62 & 1.49^{+0.09}_{-0.09} \\ 
32696 & SF & 1.625^{+0.006}_{-0.082} & 21.03 & 1.63 & 1.41 & 10.60 & 1.43^{+0.29}_{-0.18} \\ 
31086 & SF & 1.623 \tablenotemark{b} & 21.75 & 1.40 & 0.92 & 10.38 & 1.22^{+0.16}_{-0.15} \\ 
31703 & SF & 1.623 \tablenotemark{b,c} & 21.71 & 0.75 & 0.83 & 10.36 & 1.14^{+0.11}_{-0.09} \\ 
28015 & Q & 1.620^{+0.001}_{-0.001} & 21.94 & 1.61 & 0.72 & 10.30 & 0.95^{+0.44}_{-0.32} \\ 
27956 & Q & 1.629^{+0.063}_{-0.214} & 22.17 & 1.63 & 0.83 & 10.27 & 1.37^{+0.69}_{-0.42} \\ 
\hline
\multicolumn{8}{c}{\textit{mass incomplete}}\\
31715 & SF & 1.626^{+0.004}_{-0.002} & 22.24 & 0.77 & 0.82 & 10.15 & 1.23^{+0.25}_{-0.20} \\ 
30456 & SF & 1.610^{+0.028}_{-0.004} & 21.55 & 1.54 & 2.12 & 10.08 & 1.48^{+1.90}_{-0.89} \\ 
30472 & SF & 1.623 \tablenotemark{b} & 22.21 & 0.68 & 0.72 & 10.05 & 1.12^{+0.09}_{-0.10} \\ 
35210 & Q & 1.632^{+0.013}_{-0.220} & 23.11 & 1.47 & 1.01 & 9.87 & 1.17^{+1.36}_{-0.62} \\ 
29841 & SF & 1.623^{+0.001}_{-0.001} & 22.60 & 0.43 & 0.58 & 9.86 & 0.91^{+0.10}_{-0.08} \\ 
33093 & SF & 1.629 \tablenotemark{b,c} & 23.06 & 0.65 & 0.29 & 9.63 & 1.36^{+0.32}_{-0.27} \\ 
32608 & SF & 1.626 \tablenotemark{b} & 23.19 & 0.59 & 0.35 & 9.55 & 0.96^{+0.18}_{-0.17} \\ 
28036 & SF & 1.619^{+0.003}_{-0.003} & 23.24 & 0.62 & 0.07 & 9.51 & 1.30^{+0.35}_{-0.26} \\ 
29050 & SF & 1.624 \tablenotemark{b} & 23.38 & 0.75 & 0.54 & 9.48 & 0.62^{+0.16}_{-0.13} \\ 
33068 & SF & 1.624 \tablenotemark{b} & 23.60 & 0.40 & 0.11 & 9.38 & 1.12^{+0.35}_{-0.32} \\ 
30952 & SF & 1.629^{+0.001}_{-0.004} & 24.41 & 0.47 & 0.05 & 8.88 & 0.79^{+0.34}_{-0.23} \\ 
\enddata
\tablerefs{\tablenotetext{a}{\citet{Tanaka10}} \tablenotetext{b}{\citet{Tran15}} \tablenotetext{c}{\citet{Papovich10}}}
\tablecomments{Redshifts with no reported uncertainty are spectroscopic, and are averages if multiple measurements exist.}
\label{Table:cluster}
\end{deluxetable*}

\startlongtable
\begin{deluxetable*}{LcLccccR}

\tablehead{\colhead{3D-HST ID} & \colhead{Q/SF} & \colhead{$z$} & \colhead{$J_{AB_{rest}}$} & \colhead{$(U-V)_{rest}$} & \colhead{$(V-J)_{rest}$} & \colhead{$log(M/M_{\odot})$} & \colhead{$D_{n}(4000)$}}
\tablecaption{Field Galaxy Sample}
\tablenum{3}
\startdata
29179 & SF & 1.562^{+0.007}_{-0.010} & 20.25 & 1.85 & 1.71 & 11.21 & 2.48^{+2.00}_{-0.86} \\ 
32904 & Q & 1.322^{+0.001}_{-0.001} & 19.69 & 1.89 & 1.15 & 11.07 & 1.42^{+0.03}_{-0.03} \\ 
32468 & SF & 1.308^{+0.018}_{-0.030} & 20.45 & 2.23 & 2.11 & 11.00 & -0.12^{+0.51}_{-0.61} \\
34899 & SF & 1.289^{+0.001}_{-0.001} & 21.03 & 0.95 & 1.43 & 10.82 & 0.87^{+0.05}_{-0.04} \\ 
33670 & SF & 1.416^{+0.034}_{-0.005} & 20.88 & 1.87 & 1.60 & 10.65 & 2.21^{+3.79}_{-2.57} \\ 
32166 & SF & 1.324^{+0.004}_{-0.003} & 20.79 & 1.53 & 1.31 & 10.58 & 1.02^{+0.15}_{-0.13} \\ 
33472 & SF & 1.330^{+0.006}_{-0.007} & 21.24 & 1.59 & 1.24 & 10.39 & 1.59^{+0.23}_{-0.21} \\ 
28822 & Q & 1.756^{+0.002}_{-0.002} & 21.66 & 1.34 & 0.74 & 10.37 & 1.32^{+0.09}_{-0.10} \\ 
33524 & SF & 1.606^{+0.010}_{-0.233} & 22.21 & 1.58 & 1.53 & 10.35 & 1.40^{+0.92}_{-0.55} \\ 
30994 & SF & 1.278^{+0.001}_{-0.002} & 21.55 & 1.04 & 1.23 & 10.28 & 1.13^{+0.11}_{-0.18} \\ 
34916 & SF & 1.539^{+0.027}_{-0.029} & 21.93 & 1.23 & 1.27 & 10.26 & 1.51^{+0.25}_{-0.34} \\ 
27657 & SF & 1.452^{+0.007}_{-0.008} & 21.86 & 1.01 & 0.86 & 10.24 & 1.48^{+0.27}_{-0.19} \\ 
\hline
\multicolumn{8}{c}{\textit{mass incomplete}}\\
28211 & SF & 1.531^{+0.015}_{-0.014} & 22.06 & 1.36 & 1.70 & 10.14 & 16.78^{+36.52}_{-17.56} \\ 
29879 & Q & 1.670^{+0.004}_{-0.135} & 22.39 & 1.56 & 1.17 & 10.12 & 1.82^{+1.63}_{-0.91} \\ 
35978 & SF & 1.726^{+0.238}_{-0.088} & 23.10 & 1.28 & 0.93 & 10.08 & 1.76^{+3.50}_{-1.74} \\ 
32931 & Q & 1.565^{+0.004}_{-0.003} & 22.47 & 1.42 & 0.69 & 10.03 & 1.11^{+0.25}_{-0.16} \\ 
31745 & SF & 1.507^{+0.189}_{-0.050} & 22.35 & 1.27 & 0.78 & 10.01 & 1.56^{+1.37}_{-0.68} \\ 
31128 & SF & 1.553^{+0.001}_{-0.001} & 22.21 & 0.67 & 0.37 & 9.95 & 0.94^{+0.10}_{-0.07} \\ 
31079 & SF & 1.393^{+0.001}_{-0.001} & 21.87 & 0.76 & 0.71 & 9.91 & 1.11^{+0.08}_{-0.08} \\ 
36949 & SF & 1.665^{+0.005}_{-0.092} & 22.50 & 0.69 & 0.41 & 9.90 & 1.29^{+0.22}_{-0.12} \\ 
35083 & Q & 1.729^{+0.038}_{-0.028} & 23.09 & 1.41 & 0.86 & 9.89 & 1.17^{+0.96}_{-0.35} \\ 
27454 & SF & 1.409^{+0.001}_{-0.001} & 21.62 & 1.09 & 1.69 & 9.87 & 0.47^{+0.17}_{-0.16} \\ 
33889 & SF & 1.539^{+0.003}_{-0.005} & 22.46 & 0.64 & 0.52 & 9.85 & 1.05^{+0.11}_{-0.10} \\ 
28584 & SF & 1.290^{+0.001}_{-0.007} & 22.36 & 0.70 & 0.55 & 9.80 & 1.48^{+0.22}_{-0.25} \\ 
34940 & Q & 1.460^{+0.057}_{-0.022} & 22.91 & 1.38 & 0.89 & 9.80 & 1.32^{+0.49}_{-0.48} \\ 
31176 & Q & 1.554^{+0.002}_{-0.001} & 23.36 & 1.65 & 0.57 & 9.79 & 1.94^{+2.31}_{-0.93} \\ 
37588 & SF & 1.755^{+0.002}_{-0.002} & 23.10 & 0.81 & 0.43 & 9.71 & 1.12^{+0.29}_{-0.23} \\ 
27068 & SF & 1.499^{+0.002}_{-0.001} & 22.46 & 0.68 & 0.54 & 9.68 & 1.26^{+0.24}_{-0.19} \\ 
33615 & SF & 1.719^{+0.028}_{-0.047} & 23.28 & 1.13 & 0.43 & 9.66 & 1.56^{+0.60}_{-0.61} \\ 
31068 & SF & 1.718^{+0.004}_{-0.040} & 23.60 & 1.08 & 0.67 & 9.66 & 0.67^{+0.52}_{-0.31} \\ 
27102 & SF & 1.518^{+0.001}_{-0.001} & 22.98 & 0.48 & 0.78 & 9.48 & 1.56^{+0.42}_{-0.34} \\ 
35385 & SF & 1.316^{+0.009}_{-0.070} & 23.01 & 0.67 & 0.69 & 9.48 & 0.93^{+0.36}_{-0.16} \\
34110 & SF & 1.410^{+0.002}_{-0.002} & 23.27 & 0.69 & 0.50 & 9.46 & 1.20^{+0.18}_{-0.17} \\  
28706 & SF & 1.484^{+0.105}_{-0.012} & 23.38 & 0.72 & 0.25 & 9.45 & 4.57^{+5.21}_{-3.61} \\ 
33427 & SF & 1.726^{+0.001}_{-0.003} & 23.90 & 0.44 & 0.23 & 9.26 & 6.81^{+13.82}_{-10.25} \\ 
36977 & SF & 1.291^{+0.001}_{-0.002} & 23.41 & 0.65 & 0.31 & 9.25 & 0.72^{+0.15}_{-0.14} \\ 
32732 & SF & 1.695^{+0.001}_{-0.007} & 23.58 & 0.43 & 0.60 & 9.24 & 1.84^{+0.51}_{-0.44} \\ 
28114 & SF & 1.479^{+0.005}_{-0.003} & 24.32 & 0.45 & 0.11 & 9.05 & 1.28^{+1.25}_{-0.68} \\ 
\enddata
\label{Table:field}
\end{deluxetable*}

\clearpage
\onecolumngrid
\section{The 3D-HST and UltraVISTA $UVJ$ Selection Criteria}
\label{Appendix2}

\begin{figure*}[htb!]
\epsscale{1.1}
\plottwo{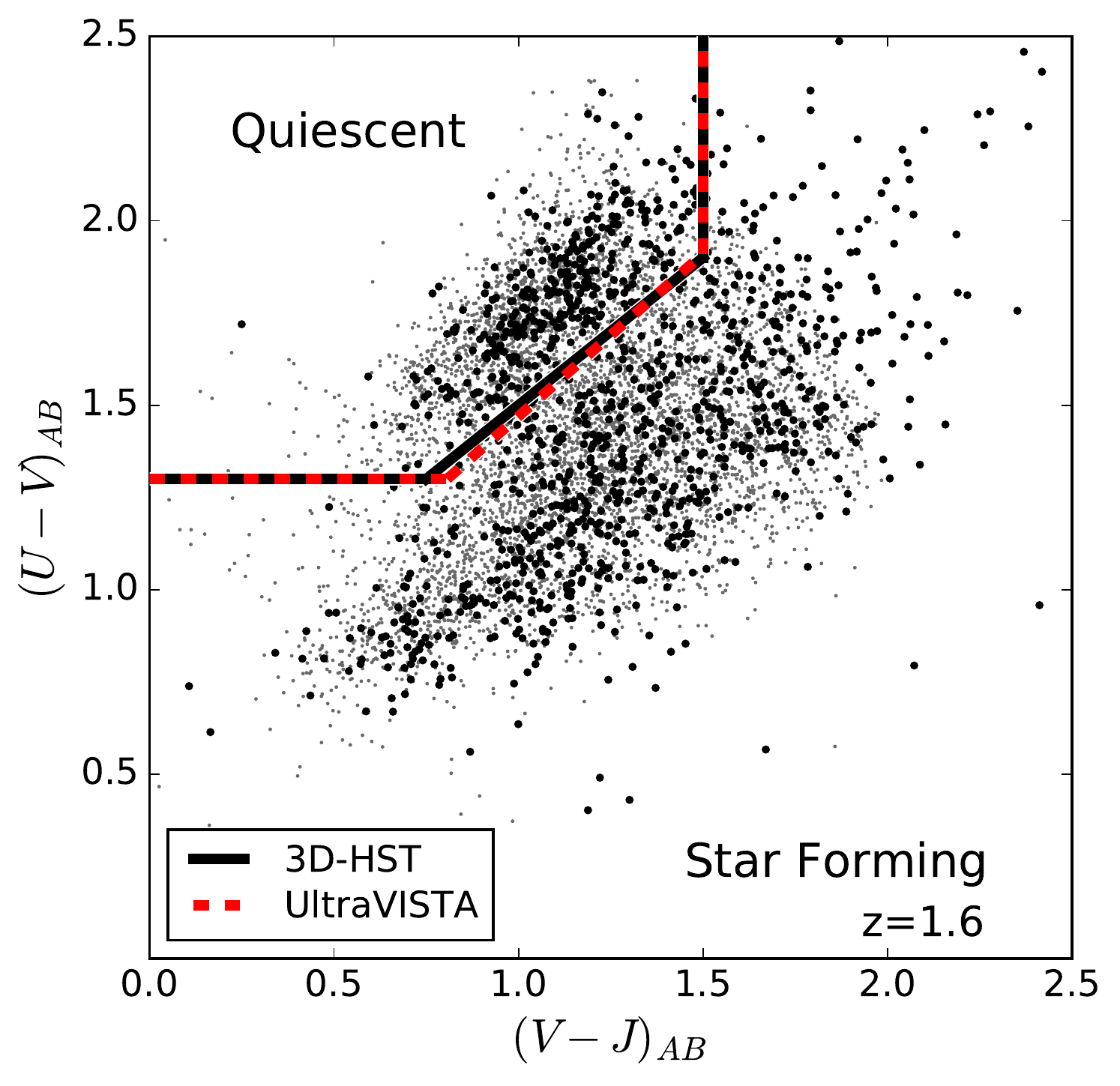}{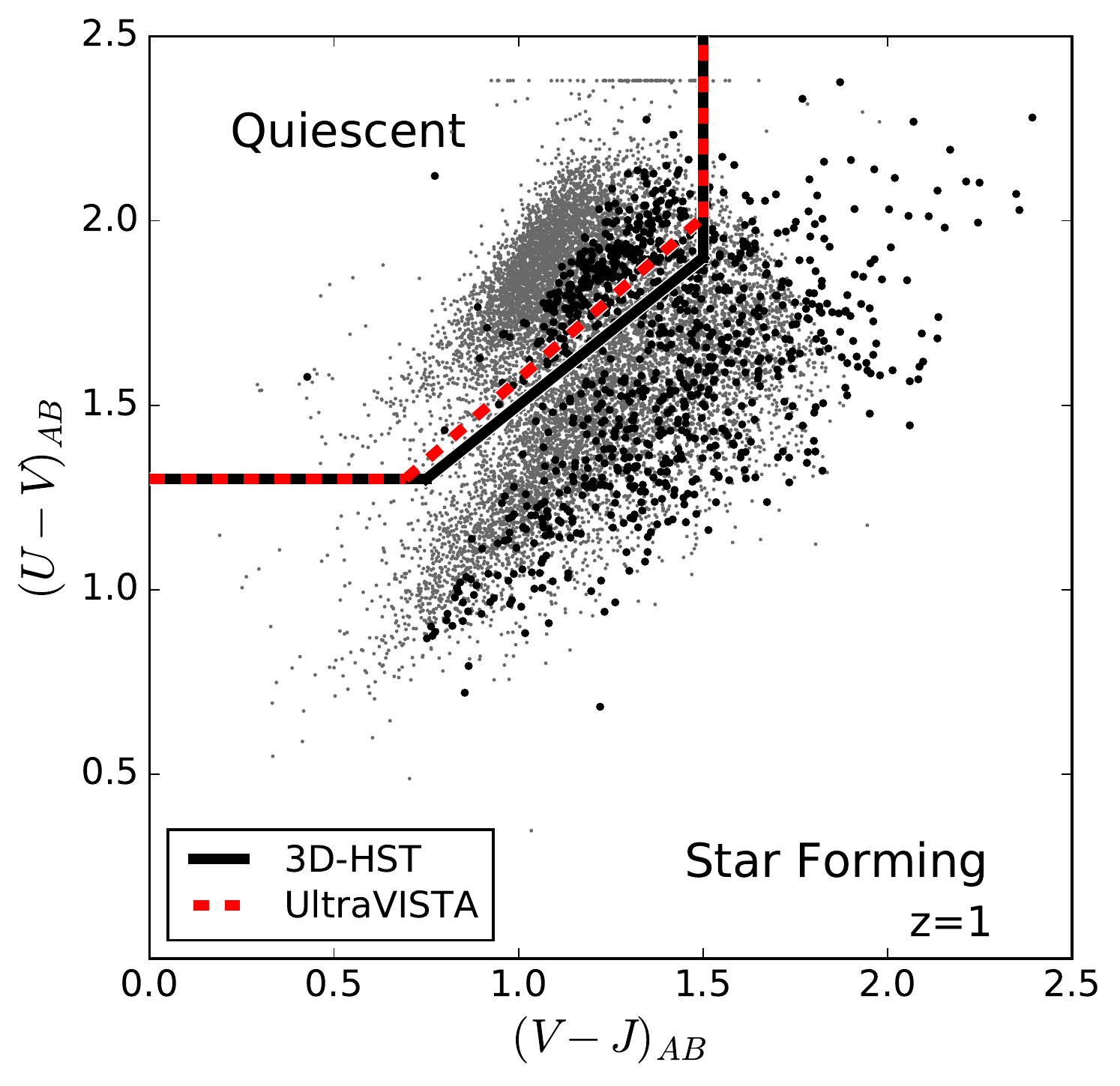}
\figcaption {Rest-frame $UVJ$ colors for $z\sim 1.6$ (left panel) and $z\sim 1$ (right) galaxies in UltraVISTA (gray background points) and 3D-HST (black foreground points). The black solid lines give the $UVJ$ selection criteria for 3D-HST galaxies as defined in \citet{Whitaker12}. The dashed red lines show the redshift-dependent criteria adopted by \citet{Muzzin13} for UltraVISTA galaxies.}
\label{Fig:uvj_selection_criteria}
\end{figure*}

In Figure~\ref{Fig:uvj_selection_criteria} we show the rest-frame $UVJ$ colors for the UltraVISTA and 3D-HST $z\sim 1.6$ and $z\sim 1$ field samples. It is apparent that an offset in rest-frame color exists between the two surveys at $z\sim 1$, but not at $z\sim 1.6$. The result of this redshift-dependent offset is that the 3D-HST $UVJ$ criteria (black lines in Figure~\ref{Fig:uvj_selection_criteria}) used by \citet{Whitaker12} adequately separate the red sequence galaxies in the 3D-HST data, but overlap with the star forming sequence of galaxies in the UltraVISTA catalog. This in turn results in a higher $f_Q$ for the UltraVISTA sample at $z\sim 1$, relative to 3D-HST. Thus, for the UltraVISTA galaxies, we instead adopt the slightly different $z \leq 1$ and $z \geq 1$ $UVJ$ criteria given in \citet{Muzzin13} (red dashed lines in Figure~\ref{Fig:uvj_selection_criteria}). These criteria better trace the gap between the red sequence and star forming galaxies in UltraVISTA at $z\sim 1$ and are nearly identical to the \citet{Whitaker12} criteria at $z\sim 1.6$. However, based on the positions of the two sets of $UVJ$ lines in Figure~\ref{Fig:uvj_selection_criteria}, it appears that even after these changes, UltraVISTA and 3D-HST galaxies are not being separated into quiescent and star forming populations in the same way at $z\sim 1$. This difference may be driving the systematic offset in $f_Q$ we observe between the two surveys at $z\sim 1$ (see Section~\ref{Sec:discussion}).

\end{document}